\documentclass[a4paper,fleqn,usenatbib]{mnras}

\usepackage{newtxtext,newtxmath}

\usepackage[T1]{fontenc}
\usepackage{ae,aecompl}
\usepackage{multicol}

\usepackage{fp}
\usepackage{xfrac}
\usepackage{hyperref}
\usepackage{booktabs}
\usepackage{cancel}
\usepackage{colortbl}
\usepackage{csquotes}
\usepackage{datatool}
\usepackage{helvet}
\usepackage{mathpazo}
\usepackage{listings}
\usepackage{multirow}
\usepackage{array}
\usepackage{float}

\usepackage{siunitx}

\usepackage{xparse}
\usepackage{etoolbox}
\usepackage{isotope}
\usepackage{ctable}
\usepackage{url}

\usepackage{graphicx}

\newcommand{\msun}{\ensuremath{\mathrm{M_{\odot}}}}

\newlength\figureheight 
\newlength\figurewidth 
\DeclareSIUnit[number-unit-product = \;]
\solarmass{\msun}

\DeclareSIUnit[number-unit-product = \;]
\dyn{dyn}

\DeclareSIUnit[number-unit-product = \;]
\ergy{erg}

\hypersetup{
    colorlinks,
    citecolor=blue,
    filecolor=black,
    linkcolor=blue,
    urlcolor=black
}

\newcommand{\RMS}[1]{{#1}_{\mathrm{RMS}}}
\newcommand{\fuelRMS}[1]{{#1}_{\mathrm{fuel,RMS}}}

\newcommand{\Lbox}{L_\mathrm{box}}
\newcommand{\Lk}{L_\mathrm{k}}

\newcommand{\Cfrac}{X_\mathrm{C}}

\newcommand{\eddyt}{\tau_\mathrm{eddy}}
\newcommand{\Dfuel}{\rho_\mathrm{fuel,i}}
\newcommand{\Dfuelavg}{\rho_\mathrm{fuel,avg}}
\newcommand{\Tfuel}{T_\mathrm{fuel}}
\newcommand{\MRMS}{\mathcal{M}_\mathrm{RMS}}

\newcommand{\urho}{\mathrm{g\,cm^{-3}}}
\newcommand{\uT}{\mathrm{K}}

\newcommand{\ut}{\mathrm{ms}}
\newcommand{\uL}{\mathrm{km}}
\newcommand{\uvkms}{\mathrm{km\,s^{-1}}}
\newcommand{\ugacc}{\mathrm{cm\,s^{-2}}}
\newcommand{\uerg}{\mathrm{erg}}

\newcommand{\norm}[1]{\left\lVert{#1}\right\rVert}

\title[Mesoscale Turbulence in SN Ia]
{Mesoscale Turbulence in Type Ia Supernova Deflagrations: Buoyancy-Driven Fuel Heating and Prospects for Delayed-Detonations}
\author[E. Brooker, A. Zhiglo, and T. Plewa]
{E. Brooker$^{1}$,
 A. Zhiglo$^{2}$,
 and T. Plewa$^{1}$\thanks{E-mail: tplewa@fsu.edu (TP)}\\
  $^1$Department of Scientific Computing, Florida State University, Tallahassee, FL 32306, U.S.A.\\
  $^2$Kharkiv Institute of Physics and Technology, Kharkiv, 61108, Ukraine, on leave from
}
\pubyear{2025}
\begin{document}
\pagerange{{000}--{000}} \pubyear{2020} \volume{000}
\label{firstpage}
\maketitle
%
%
%
%
%
%
\begin{abstract}
%
%
The aim of this work is to characterize the thermodynamic state of fuel mixed into the turbulent flame brush in the context of the Zel'dovich deflagration-to-detonation transition (ZDDT) mechanism of Type Ia supernovae (SNe Ia).
We perform a series of three-dimensional computer simulations of thermonuclear deflagrations subject to the Rayleigh-Taylor instability (RTI) for conditions found in model explosions of centrally ignited realistic, Chandrasekhar mass white dwarf progenitors. These conditions correspond to explosion times when the flame reaches low density progenitor regions where DDT is expected to occur.
The flame database is constructed using a thickened flame model. High numerical resolution is achieved with the help of the adaptive mesh refinement (AMR) approach allowing, for the first time, to resolve mesoscale buoyancy-driven flame turbulence. The system is evolved to a quasi-steady state, and flow properties in the turbulent region, where turbulence is most isotropic, is analyzed in a co-moving frame of reference.
We find evidence for strong buoyancy-driven adiabatic heating of fuel layers adjacent to the flame front. The heating results in a dramatic reduction of fuel ignition times by between $\approx$2 and more than about 5 orders of magnitude. The heating increases with the RTI forcing.
The observed shortening of fuel burning timescales suggests a new source of energy is important inside fuel penetrating the flame brush. These regions are up to several hundred meters wide. On the basis of the previous results of turbulent combustion in SNe Ia, preconditioning required by the ZDDT mechanism can occur there.
\end{abstract}
%
%
\begin{keywords}
stars: white dwarfs --- supernovae: general --- hydrodynamics, instabilities, turbulence --- nuclear reactions
\end{keywords}
%
%
%
%
%
%
\section{Introduction}\label{s:intro}
The DDT mechanism remains one of the major unsolved problems in theoretical and computational combustion. It has been directly observed in a number of laboratory experiments and extensively studied by means of computer simulations \citep[][and references therein]{Oran2007}. In the context of astrophysics and stellar evolution, it has now been suspected for over 30 years that DDT is directly responsible for at least a subclass of white dwarf explosions producing luminous SNe Ia. This evidence for DDT is indirect, relying on post-explosion observational data and phenomenological models such as delayed-detonation \citep[DD;][]{Hoeflich1995}. For these DD models to match observations, DDT must occur at relatively low densities, $\varrho_\mathrm{tr}\approx 2\times 10^7$ g cm$^{-3}$. Subsequently, this conclusion was quickly linked to a morphological change from the flamelet to stirred flame burning regime \citep[FSF transition;][]{Niemeyer+97,Khokhlov1997b} expected of centrally-ignited deflagrations. As theorized by \cite{Khokhlov1991b}, DDT might be possible provided that turbulent perturbations (Rayleigh-Taylor driven and flame generated) satisfy certain minimum amplitude and size requirements.

The above basic DD explosion scenario was refined in the course of major research efforts. \cite{Woosley+09} used a one-dimensional model of a turbulent flame to study the flame evolution in various burning regimes, including the FSF transition. They provided firmer constraints on conditions for DDT, speculated about possible collective effects in multidimensions, and in follow-on work \citep{Woosley+11} assessed the role of fuel composition. \cite{Schmidt+10} adopted a probabilistic DDT framework originally developed by \cite{Pan+08} to account for intermittency of turbulence as a possible source of large-scale fluctuations at the flame front. They provided an independent set of DDT constraints and discussed their dependence on the adopted intermittency model. More recently \citet{Poludnenko2019} (see, also, \citet{Gordon2021}) postulated a similarity between the terrestrial and astrophysical DDT mechanisms. In that scheme, interaction between a deflagration with nearly incompressible turbulence produces a shock and ultimately results in a detonation. However, neither this nor other briefly discussed above efforts demonstrated DDT to operate under conditions expected for SNe Ia.

In our previous work \citep[][BFP21 below]{Brooker2021}, we studied weakly compressible ($\MRMS\approx0.37$) turbulence models with nuclear burning effects to identify the conditions necessary for DDT to occur in SN Ia explosions. We observed the formation of carbon deflagrations and their transitions to carbon detonations due to the Zel'dovich mechanism \citep{Zeldovich1970}. The transition mechanism involved fuel confined in channels bounded by flame fronts and compressed at intermediate turbulent scales. This compression lead to formation of suitably preconditioned regions of fuel such that sustained carbon detonations were initiated along reactivity gradients. While these models captured the effects of turbulence and thermonuclear self-heating, carbon deflagrations were modeled implicitly via numerical diffusion.

In this work, we extend our previous studies \citep{Brooker2021} with modeling of flame-fuel interactions in Rayleigh-Taylor-driven turbulence to characterize the conditions in the fuel during the early carbon deflagration phase of SN Ia. The characterization of the fuel conditions during this phase of evolution is necessary to determine the viability of the Zel'dovich mechanism previously observed in our turbulence models.

%
%
%
\section{Methods and initial conditions}\label{s:methods}
We performed computer simulations of the RT-unstable flames using {\sc Proteus}, a custom development version of the {\sc FLASH} code \citep{Fryxell+00}. We used the USM unsplit solver \citep{LeeDeane2009} to advance in time the set of equations describing hydrodynamic evolution of stellar plasma:
\begin{align*}
 \frac{\partial \rho}{\partial t} + \nabla \cdot (\rho \mathbf{v}) &= 0,  \\
 \frac{\partial \rho \mathbf{v}}{\partial t} + \nabla \cdot (\rho \mathbf{v} \otimes \mathbf{v}) + \nabla P &= -\rho\mathbf{g}, \\
 \frac{\partial \rho E}{\partial t} + \nabla \cdot [\mathbf{v}(\rho E + P)] &= -\rho\mathbf{v}\cdot\mathbf{g} + \dot{Q}_\mathrm{E,nuc}, \\
 \frac{\partial \rho\mathbf{X}}{\partial t} + \nabla \cdot (\rho\mathbf{X} \mathbf{v}) &= \dot{Q}_\mathrm{X,nuc},
\end{align*}
where $\rho$ is the mass density, $\mathbf{v}$ is the velocity, $E$ is the specific total energy, $P$ is the pressure, $\mathbf{X}$ are mass fractions of representative nuclear species \citep{timmes+00}, and $\mathbf{g}$ is a gravitational acceleration, assumed constant in our model, $\mathbf{g}=[-g,0,0]^\mathrm{T}$. The thermodynamic state of the stellar material was described using the Helmholtz equation of state \citep{TimmesSwesty2000}.

The above set of equations was supplemented by an equation describing diffusion of nuclear species, and two local source terms. The first source term, $\dot{Q}_\mathrm{X,n}$, describes time evolution of species mass fractions, while the second source term, $\dot{Q}_\mathrm{E,n}$, accounts for the nuclear energy generation due to a subset of dominant nuclear reactions. These additional equations constitute an advection-diffusion-reaction (ADR) solver \citep{Khokhlov1991a, Khokhlov1995}, which captures time evolution of a thermonuclear deflagration in a thickened flame-type approximation \citep{colin+00}. Of particular note is the arrangement of nuclear species into three groups, which are related three main stages of nuclear kinetics of a carbon-oxygen mixture \citep[3-stage flame model; see Section 2.2 in][]{Khokhlov1991a}. ADR evolution of those groups is tracked with the help of corresponding flame progress variables. Those progress variables are subjected to ADR with their diffusion rates set to produce a desired model flame speed and reactions-related terms approximately accounting for transformations of nuclear species and the flame energetics. The first of those variables, $\phi$, represents carbon, and is commonly used, including this work, to identify the flame front and analyze its dynamics.

The original Khokhlov ADR scheme was improved to correct the flame profile for thermal expansion effects \citep{Zhiglo07} and reduce its sensitivity to strain and stretch, as described in Model B of \citet[][Appendix A]{Zhiglo09}. The latter improvement made the flame profile more uniform in terms of its structure; undesirable exponential tails of (the first) progress variable present in the original model were also eliminated. For detailed discussion of the flame model performance we refer the interested reader to \citet{Zhiglo09}.

Simulations were performed on a 3-dimensional domain in Cartesian geometry. The domain was $768\ \uL$ long in the longitudinal, gravity-aligned direction (along the x-axis) and $32\ \uL$ long in both lateral directions. We used a reflecting and outflow boundary conditions at lower and upper boundary in the longitudinal direction, and periodic boundary conditions on the lateral boundaries. Although we observe newly born RTI bubbles continuously generating weak acoustic perturbations that propagate into the fuel, we have not observed any significant reflections of those perturbations from the upstream boundary that could potentially affect the interior solution.

The mesh was dynamically adapted throughout the simulation with the refinement conditions examined every other step. The base mesh level contained 24 blocks, each block with 16 mesh cells per dimension, arranged along the longitudinal direction. We allowed for additional five levels of refinement, with the resolution between the consecutive levels increasing by a factor of two. At the finest mesh level, the effective mesh resolution was 12,288 cells in the longitudinal direction and 512 cells in the lateral directions. This corresponds to an effective mesh resolution of $\Delta x_\mathrm{eff} = 62.5$\,m in each dimension. Starting at the base mesh level, the mesh resolution was increased in regions where any of the three ADR flame progress variables varied between 0.01 and 0.99. Furthermore, in order to prevent refinement-induced perturbations, we have found it necessary also to refine regions where pressure degeneracy had been lifted, $T > 3\times10^{9}$ K, and local relative pressure variations met the condition, $\delta p/p > 1\times10^{-3}$. To additionally stabilize the mesh blocks were assigned a finite minimum lifetime of 5 time steps.

The summary of model flame initial conditions is given in Table \ref{t:modelICs}.
\begin{table*}
    \caption{Model flame initial conditions.}
    \label{t:modelICs}
    \begin{center}        
    \begin{tabular}
    {l c c c c c c c c c}
    \FL
    model & $\Dfuel$\tmark[a]  &  $g$\tmark[b]  &  $\Tfuel$\tmark[c]  &  $\tau_\mathrm{ign}$\tmark[d] &  $\Cfrac$\tmark[e]  &  $v_\mathrm{f,SGS}$\tmark[f]  &  $t_{L512}$\tmark[g]  &  $t_\mathrm{i}$\tmark[h]  &  $t_\mathrm{f}$\tmark[i] \NN
          & $\urho$            &  $\ugacc$      &  $\uT$              &  s                            &                     &  $\uvkms$                     &  s                    &  s                        &  s                       \ML
    LOD   & \num{6e6}          &  \num{1.1e9}   &  \num{1.5e9}        &  \num{0.33}                    &  \num{0.420}        &  \num{8.96}                   &  1.200                &  1.480                    &  1.688                   \NN
    MED   & \num{2e7}          &  \num{2.5e9}   &  \num{1.3e9}        &  \num{0.33}                    &  \num{0.425}        &  \num{12.1}                   &  1.000                &  1.100                    &  1.243                   \NN
    HID   & \num{1e8}          &  \num{6.4e9}   &  \num{1.2e9}        &  \num{0.34}                    &  \num{0.330}        &  \num{15.2}                   &  0.612                &  0.650                    &  0.708                   \LL

    \multicolumn{10}{l}{$^a$ Initial fuel density.}\\
    \multicolumn{10}{l}{$^b$ Gravitational acceleration.}\\
    \multicolumn{10}{l}{$^c$ Fuel temperature.}\\
    \multicolumn{10}{l}{$^d$ Fuel ignition time.}\\
    \multicolumn{10}{l}{$^e$ Fuel carbon abundance.}\\
    \multicolumn{10}{l}{$^f$ RTI-based ADR subgrid-scale flame speed.}\\
    \multicolumn{10}{l}{$^g$ Simulation time at which the effective mesh resolution reached 512 mesh cells laterally.}\\
    \multicolumn{10}{l}{$^h$ Initial simulation time of the analyzed time interval.}\\
    \multicolumn{10}{l}{$^i$ Final simulation time of the analyzed time interval.} 
    \end{tabular}
    \end{center}
\end{table*}
We explored flame evolution at three initial fuel densities covering the density range thought to harbor conditions amenable to the ZDDT mechanism \citep{Hoeflich1995, Woosley+07, Pan+08}. In what follows, these models are identified as LOD, MED, and HID, respectively, and collectively referenced to as Rayleigh-Taylor-driven (RTD) models.

The table also provides gravitational acceleration and chemical composition of fuel at the selected initial densities, which were adopted from a two-dimensional pure deflagration SN Ia model for conditions found when the flame reaches the target densities. The explosion model used a realistic progenitor structure from a low-mass close binary evolution model \citep{WongSchwab2019} with a centrally ignited flame and the ignition conditions matching the \textsc{n11r100y00} models series of \citet{Plewa2007}. The explosion simulation was performed using a customized version of the \textsc{FLASH} code \cite{Plewa2007}, but with the improved ADR flame, as described earlier in this section.

We adjusted the initial fuel temperature found in the explosion model so that the burning timescale of the fuel was slightly longer than 300\,ms (see Table \ref{t:modelICs}). This amount of time on the first hand is sufficiently long for RTI to fully develop turbulence in our models (had a self-heating be included), and on the second hand is still available for that material before the flame enters the outermost layers of the progenitor and quenches in the explosion models. During our simulations, the fuel density gradually was slowly decreasing from its initial value during the course of the simulation due to expansion of the fuel region. The average fuel densities just upstream of the flame front at the time we make our analyses for each model, and which should be considered as actual model RT flame densities when compared to similar studies, are given in Table \ref{t:modelAnalyzed}.
\begin{table}
    \caption{Model flame properties during the analyzed time interval.}
    \label{t:modelAnalyzed}
    \begin{center}        
    \begin{tabular}
    {l c c c c c }
    \FL
    model  &  $\Dfuelavg$\tmark[a]  &  $v_\mathrm{f}$\tmark[b]  &  $\hat{S}_\mathrm{f}$\tmark[c]  &  $\tau_\mathrm{ign,min}$\tmark[d]  &  $\tau_\mathrm{ign,max}$\tmark[e]  \NN
           &  $\urho$               &  $\uvkms$                 &                                 &  s                                 &  s                                 \ML
    LOD    &  \num{5.0e6}           &  \num{163}$\pm$\num{19}   &  \num{17.2}$\pm$\num{0.2}       &  \num{6.4e-4}                      &  \num{4.0e-3}                      \NN
    MED    &  \num{1.4e7}           &  \num{247}$\pm$\num{25}   &  \num{18.9}$\pm$\num{0.4}       &  \num{1.6e-5}                      &  \num{2.7e-4}                      \NN
    HID    &  \num{6.3e7}           &  \num{475}$\pm$\num{69}   &  \num{23.4}$\pm$\num{3.5}       &  \num{1.6e-8}                      &  \num{4.9e-7}                      \LL

    \multicolumn{6}{l}{$^a$ Average fuel density.}\\
    \multicolumn{6}{l}{$^b$ Turbulent flame speed.}\\
    \multicolumn{6}{l}{$^c$ Flame surface area increase factor.}\\
    \multicolumn{6}{l}{$^d$ Minimum fuel ignition time at $\phi=\num{1e-3}$.}\\
    \multicolumn{6}{l}{$^e$ Maximum fuel ignition time at $\phi=\num{1e-3}$.}
    \end{tabular}
    \end{center}
\end{table}

The adopted in this work explosion model-based initial conditions differ from those used in earlier similar RTI flame studies by \cite{Khokhlov1995} and \cite{Zhang2007}. Those earlier works focused on scaling properties of specific flame characteristics, such as the turbulent flame speed, in quasi-steady state. Those properties of interest were evaluated and their behavior probed around a single reference state with fuel density ($1\times10^8$ g cm$^{-3}$), gravitational acceleration ($1.9\times10^{9}$ cm s$^{-2}$), and composition (equal mass fractions of carbon and oxygen). That reference state density matches that of our HID model, but differs in terms of both gravitational acceleration and fuel composition.

The ADR flame model uses a predefined constant during the evolution subgrid-scale flame speed, $v_\mathrm{f,SGS}$, based on the RTI growth rate. The growth rate itself depends on the gravitational acceleration and the flame front Atwood number, which vary between models. In our models, RTI was seeded by imposing a sinusoidal perturbation at the flame front with amplitude of $1$ km and three wavelengths per lateral direction and the nominal flame location, $x_\mathrm{f,i}$, equal to $25$ km.

Each series of simulations were initiated with low resolution (128 mesh cells laterally) runs. Those runs were continued until the flame evolution reached a quasi-steady state, at which time the effective mesh resolution was doubled. The model mapping process used conservative interpolation provided by the AMR package with the third-order accuracy in space. It is conceivable that this initial mapped state had to be evolved for some time to allow the just-mapped solution to adjust to the increased mesh resolution. At the time when the new quasi-steady state was reached, denoted $t_\mathrm{L512}$ in Table \ref{t:modelICs}, the mesh resolution was doubled again, now to reach 512 mesh cells laterally and the effective mesh resolution of about 62 meters.

In the following analysis we disregarded the results obtained during at least the first 3\,$\eddyt$ of the high resolution model runs to allow the solution to adjust to the new mesh. This solution-to-resolution adjustment period of time is given by $t_\mathrm{i} - t_\mathrm{512}$, as shown in Table \ref{t:modelICs}. After that period, the models were evolved further in time in quasi-steady state for about 5\,$\eddyt$ (see $N_\tau$ in Table \ref{t:turbTPF}).
\begin{table*}
    %
    \caption{Select properties of the model TPF turbulence in high resolution models.}
    \label{t:turbTPF}
    \begin{tabular}{l c c c c c c c c}
    \FL
    model & $\Lk$\tmark[a]  & $k_\mathrm{c}$\tmark[b]  & $k_\mathrm{max}$\tmark[c]  &  $\eddyt$\tmark[d]     & $N_{\tau}$\tmark[e]  & $\bar{E}_\mathrm{k}$\tmark[f]   & $R_\mathrm{xx,max}$\tmark[g]  &  $(R_\mathrm{yy}+R_\mathrm{zz})_\mathrm{max}$\tmark[h]  \NN
          & $\Lbox$         & $ $                      & $ $                        &  $\ut$                 & $ $                  & $\uerg$                         & g\,cm$^{-1}$s$^{-2}$          &  g\,cm$^{-1}$s$^{-2}$                                   \ML
    LOD   & \num{0.83}      & \num{0.54}               & \num{2.18}                 &  $\num{44}\pm\num{4}$  & \num{4.7}            & \num{4.20e+20}                  & \num{1.4e21}                  &  \num{1.3e21}                                           \NN
    MED   & \num{0.81}      & \num{0.45}               & \num{1.78}                 &  $\num{25}\pm\num{3}$  & \num{5.7}            & \num{3.98e+21}                  & \num{8.2e21}                  &  \num{5.7e21}                                           \NN
    HID   & \num{0.70}      & \num{0.27}               & \num{1.08}                 &  $\num{11}\pm\num{1}$  & \num{5.2}            & \num{3.08e+22}                  & \num{1.5e23}                  &  \num{1.1e23}                                           \LL
    \multicolumn{9}{l}{$^a$ Integral length scale.}\\
    \multicolumn{9}{l}{$^b$ RTI critical mode.}\\
    \multicolumn{9}{l}{$^c$ RTI maximum unstable mode.}\\
    \multicolumn{9}{l}{$^d$ Turbulence eddy turnover time on the integral length scale.}\\
    \multicolumn{9}{l}{$^e$ Number of analyzed eddy turnover times.}\\
    \multicolumn{9}{l}{$^f$ Time-averaged turbulent kinetic energy.}\\
    \multicolumn{9}{l}{$^g$ Maximum of the radial Reynolds stress tensor component.}\\
    \multicolumn{9}{l}{$^h$ Maximum of the sum of lateral Reynolds stress tensor components.}    
    \end{tabular}
\end{table*}
Simulations were continued at that highest effective resolution for additional about $N_\tau\,\eddyt$, where $\eddyt$ is the eddy turnover time (see Table \ref{t:turbTPF}). The actual time span of models used for the analysis was on average $\approx 5\eddyt$.

Despite the adopted gradual model resolution increase and with the AMR capability allowing us to reduce the mesh filling factor to about 15 per cent, each model proved relatively demanding computationally, requiring about two million CPU hours of computing time on the DoE NERSC Cray \textsc{Perlmutter} Cray EX supercomputer. The model output amounted to several 10 TB of data.
\section{Results}\label{s:results}
As we mentioned in the previous section, our high-resolution models were initialized with data obtained from a sequence of low-resolution runs in order to save computational time. In Table \ref{t:modelICs} we provide the final simulation times that each model was evolved to, $t_\mathrm{f}$, noting that the lower the initial flame density, $\rho_\mathrm{fuel,i}$, the longer simulation time required to reach the quasi-steady state on the finest grid resolution. This behavior is expected due to the flame speed decreasing with density as prescribed by the subgrid-scale flame capturing methodology. Once the simulations performed at the highest resolution reached their quasi-steady states, we began investigating the morphological properties of the flame solutions and the flame-generated turbulence. These model characteristics are presented next.
\subsection{The turbulent post-flame region}\label{s:tpf}
To enable a meaningful comparison of our current results with those previously reported in \citet{Brooker2021}, it is necessary to identify the TPF region. This region is a part of the post-flame region in the RT-driven models where turbulence properties most closely match those of isotropic turbulence (which is produced in the SD models). With TPF defined, we can proceed to obtain statistical measures characterizing the RT-driven turbulence and evaluate them in the context of SD turbulence (for in-depth analysis of SD turbulence models, please see \citet{Benzi2010} and also \citet{FennPlewa2017} for the stellar conditions).

\begin{figure*}
    \begin{center}
    \includegraphics[width=0.95\linewidth]{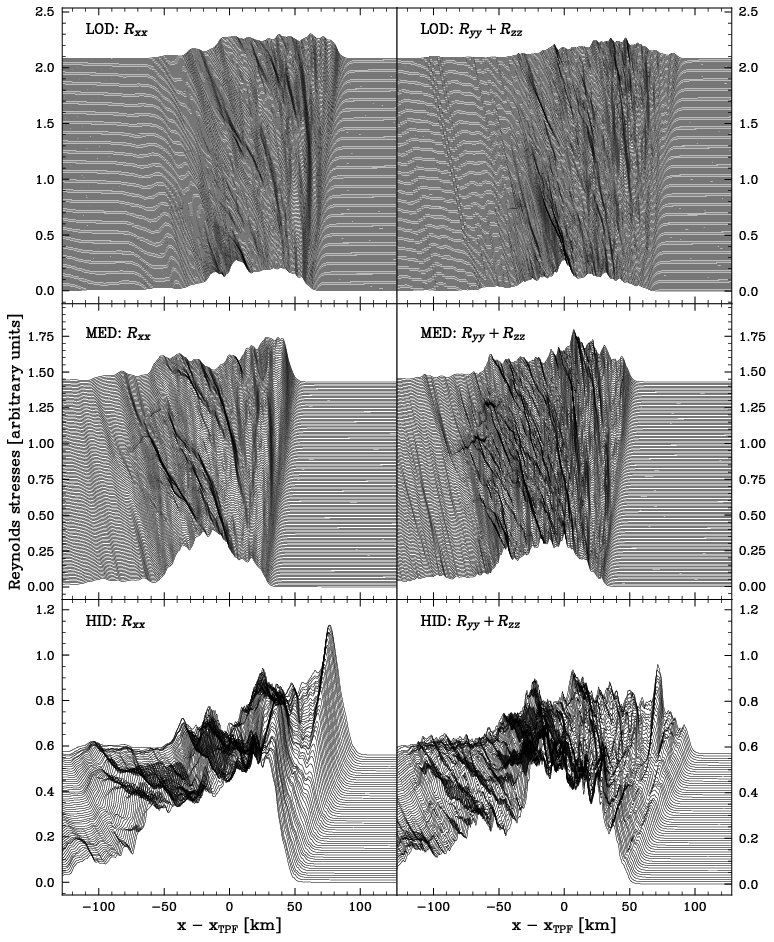} 
    \caption{The Reynolds stress tensor components are shown in a reference frame co-moving with the TPF region for all analyzed time slices for the LOD (top panel), MED (middle panel), and HID (bottom panel) models. The laterally-averaged radial (aligned with direction of gravity), $R_\mathrm{xx}$, and the sum of the lateral, $R_\mathrm{yy} + R_\mathrm{zz}$, Reynolds stress tensor components are shown in the left and right columns, respectively. For clarity of presentation of time evolution, the averages were first scaled by $(2t_\mathrm{f}R_\mathrm{max})^{-1}$ , where $t_{f}$ is the final simulation time and individual $R_\mathrm{max}$ values are given in Table \ref{t:turbTPF}. The averages for individual time slices were shifted by $10\times (t - t_{i})$ offset, where $t_{i}$ is the initial time of the model, and the so transformed time was normalized by the eddy turnover time, $\eddyt$. See text for discussion.}
    \label{f:rs}
    \end{center}
\end{figure*}

Our method of identifying a post-flame region in which turbulence becomes approximately isotropic is based on comparison of the Reynolds stress tensor components, which in isotropic turbulence should be of comparable magnitude. The time evolution of the longitudinal component (aligned with the direction of gravity) and of the sum of the lateral components of the Reynolds stress tensor are shown in a reference frame co-moving with the flame brush in the left and right columns of Fig.\ \ref{f:rs},
respectively.

In each panel of the figure, a set of lines show the laterally averaged values of the corresponding stress tensor components starting with the bottom curve, which corresponds to time $t_\mathrm{i}$, the remaining curves are offset by a constant amount with the last, top curve corresponding to time $t_\mathrm{f}$. The flat segment of the curves for $x-x_\mathrm{TPF} > 50$ km corresponds to the unperturbed fuel upstream of the flame. This feature allows to easily spot the leading edge of the evolving flame as indicated by the rapid increase in magnitude of the Reynolds stress components, starting at $x-x_\mathrm{TPF} \approx 50$ km, is associated with the leading edge of the flame front with turbulence being rapidly generated by RTI. The contribution of individual RT bubbles to the stress generation can be easily seen in the form of isolated maxima that grow and decay in amplitude in time (starting from the bottom to the top of each panel). In our models, the flame brush region extends for about 100 km with all tensor stress components gradually decreasing in value throughout this region as the fuel is consumed and the corresponding amount of buoyant energy available for driving decreases. 

\begin{figure*}
    \begin{center}
    \includegraphics[width=\linewidth]{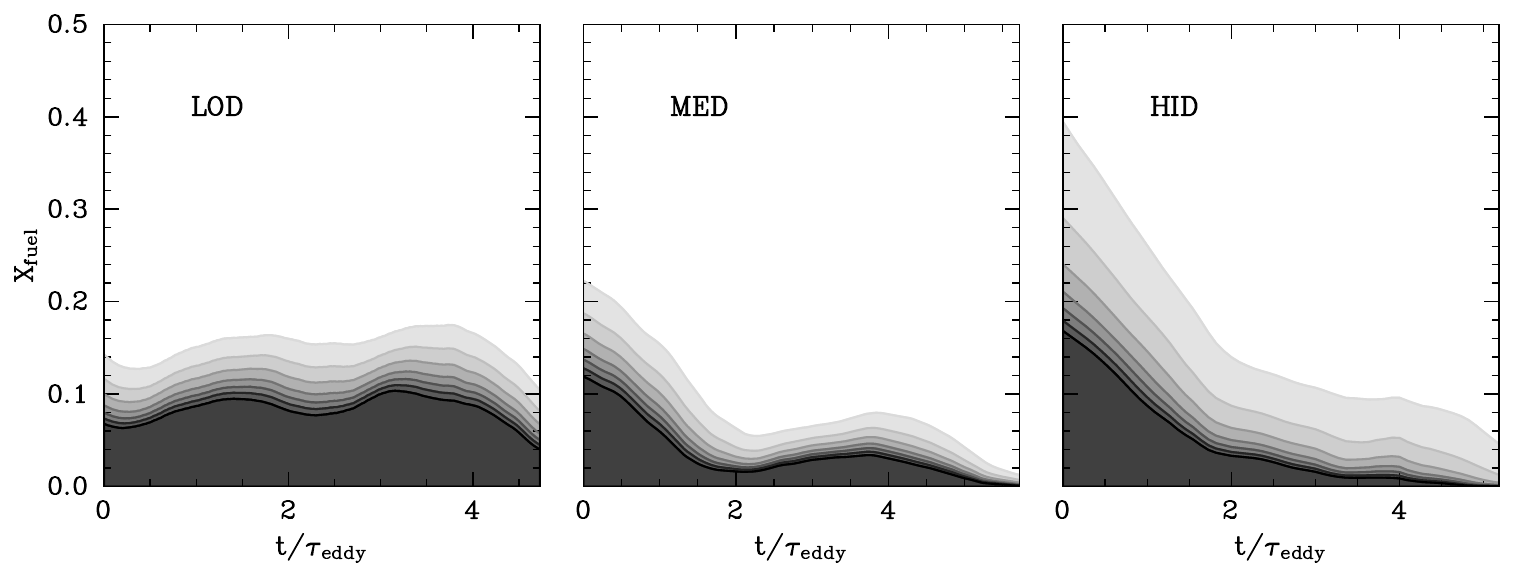}
    \caption{Flooded-contour plot of the fuel mass fraction, $X_\mathrm{fuel}$, defined in terms of the flame progress variable value, in the TPF region as a function of time for the LOD (left panel), MED (middle panel), and HID (right panel) model. The flooded-contours shaded from darkest gray to lightest gray correspond to maximum values of the flame progress variable of $1\times10^{-5}$, $1\times10^{-3}$, 0.01, 0.1, 0.3, and 0.999. Note that the horizontal axis has different spacing with the maximum corresponding the final analyzed time (cf.\ Table \ref{t:turbTPF}) in each panel. See text for details.}
    \label{f:fuelfrac}
    \end{center}
\end{figure*}

We identify the TPF region using the following procedure. First, we estimate the average longitudinal position of the flame brush by approximating sum of lateral averages of Reynold stress tensor components using a super-Gaussian profile,
\begin{equation}    
   f(A,x,x_{0}) = A e^{-2 ((x-x_{0}) \sigma^{-1})^{p}},
\end{equation}
where $A$ is a scaling factor, $x$ is the radial coordinate, $x_0$ is the center of the profile, and we use $p=10$. The fit is obtained for all model times between $t_\mathrm{i}$ and $t_\mathrm{f}$ and performed with the help of a nonlinear least squares method as implemented in the \textsc{scipy.optimize.curve\_fit} procedure \citep{scipy2020}. This set of instantaneous TPF region locations is next adjusted such that over the analyzed time period it contains a small amount of fuel (see Fig.\ \ref{f:fuelfrac}
%
%
for the time evolution of the amount of fuel inside the TPF region). The TPF centroid is finally identified as the middle of the 32 km wide region that moves with the average flame speed, $v_\mathrm{f}$, as given in Table \ref{t:flameTPF},
\begin{table*}
    \caption{Select time-average flame and TPF region properties.}
    \label{t:flameTPF}
    \begin{tabular}
    {l c c c c c c c c }
    \FL
    model & $v_\mathrm{8}$\tmark[a] & $\RMS{v}$\tmark[b]      &  $\RMS{M}$\tmark[c]           &  $\chi_\mathrm{v}$\tmark[d]  &  $\fuelRMS{v}$\tmark[e]    &  $\fuelRMS{M}$\tmark[f]      & $\chi_\mathrm{v,fuel}$\tmark[g] & $\bar{X}_\mathrm{fuel}$\tmark[h]   \NN
          & $\uvkms$                & $\uvkms$                &                               &                              &  $\uvkms$                  &                              &                                 &                                    \ML 
    LOD   & 63                      & $\num{200}\pm\num{4}\ $ &  $\num{0.044}\pm\num{0.001}$  &  $\num{1.18}\pm\num{0.03}$   &  $\ \num{304}\pm\num{29}$  &  $\num{0.080}\pm\num{0.008}$ & $\num{2.6}\pm\num{0.2}$         & \num{0.053}$\pm$\num{0.012}        \NN
    MED   & 86                      & $\num{276}\pm\num{23}$  &  $\num{0.056}\pm\num{0.005}$  &  $\num{1.20}\pm\num{0.05}$   &  $\ \num{519}\pm\num{59}$  &  $\num{0.124}\pm\num{0.014}$ & $\num{2.8}\pm\num{0.6}$         & \num{0.017}$\pm$\num{0.015}        \NN
    HID   & 205                     & $\num{631}\pm\num{71}$  &  $\num{0.079}\pm\num{0.012}$  &  $\num{1.07}\pm\num{0.18}$   &  $\num{1003}\pm\num{95}$   &  $\num{0.194}\pm\num{0.019}$ & $\num{2.6}\pm\num{0.9}$         & \num{0.010}$\pm$\num{0.014}        \LL
    \multicolumn{9}{l}{$^b$ Turbulent RMS velocity on a scale of 8 km.}\\
    \multicolumn{9}{l}{$^b$ Turbulent RMS velocity.}\\
    \multicolumn{9}{l}{$^c$ Turbulence Mach number.}\\
    \multicolumn{9}{l}{$^d$ Turbulent velocity anisotropy.}\\
    \multicolumn{9}{l}{$^e$ Fuel turbulent RMS velocity.}\\
    \multicolumn{9}{l}{$^f$ Fuel turbulence Mach number.}\\
    \multicolumn{9}{l}{$^g$ Fuel turbulent velocity anisotropy.}\\
    \multicolumn{9}{l}{$^h$ Fuel mass fraction.}  
    \end{tabular}
\end{table*}
and simultaneously satisfies the two previously specified criteria. To summarize, geometrically the TPF region is a cube with 32 km per side that contains a residual amount of fuel (see Fig.\ \ref{f:fuelfrac} and Table \ref{t:flameTPF}), and inside which lateral turbulent motions, expressed in terms of the Reynolds stresses, are near their maximal values (cf.\ Table \ref{t:turbTPF} and Fig.\ \ref{f:rs}).

We also calculated radial profiles of the Reynolds stress tensor, reporting in Table \ref{t:turbTPF} the maximum value found in each model for both the longitudinal tensor component, $R_\mathrm{xx}$, and the sum of the lateral tensor components, $R_\mathrm{yy}+R_\mathrm{zz}$. A useful indicator of turbulent-driving, we see that the maximum of these two values increases with increasing density. We note that, except for the lowest density model, the longitudinal component of the stress tensor is larger than the sum of the lateral components. This is indicative of the buoyancy forces along the longitudinal axis driving turbulence in these models.
\subsection{General flame characteristics}\label{disc_genflamchar}
\begin{figure*}
    \begin{center}
    \begin{tabular}{ccc}
        \includegraphics[width=0.64\columnwidth]{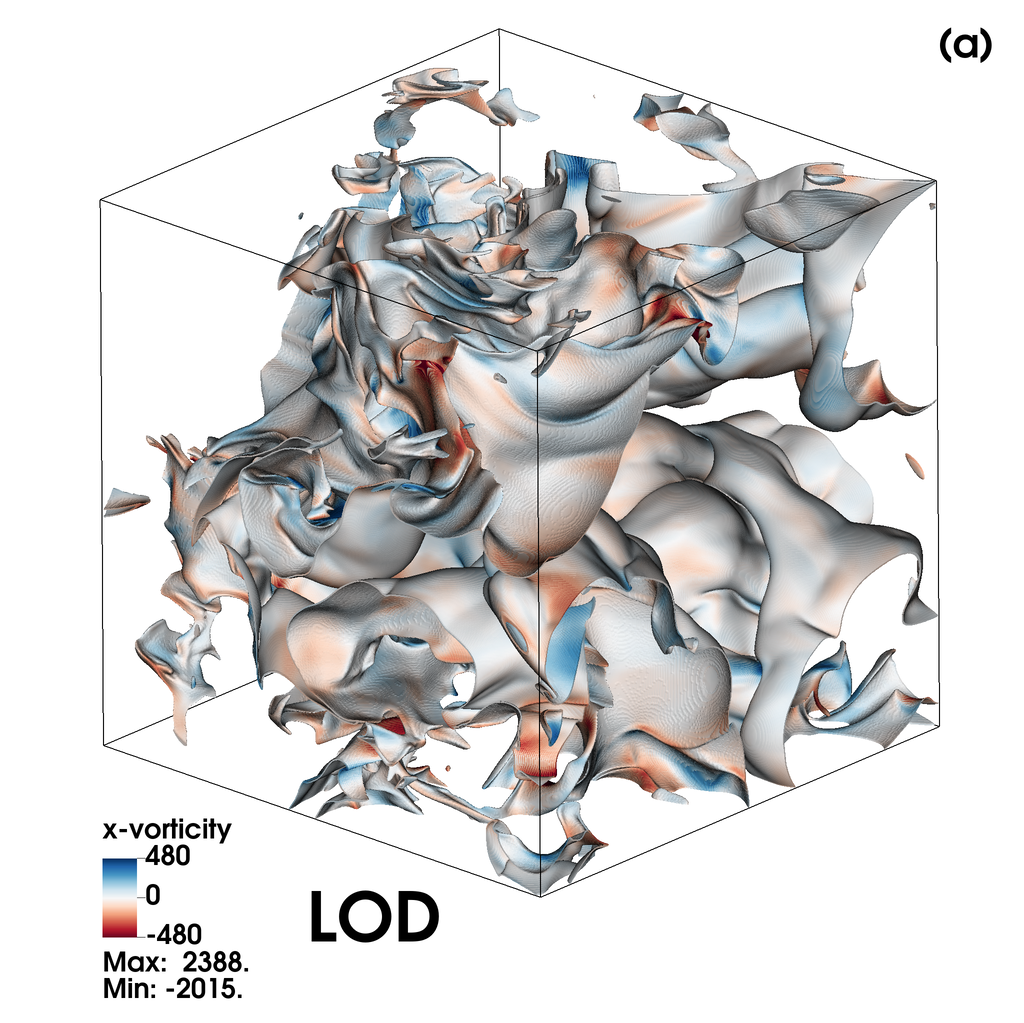} &
        \includegraphics[width=0.64\columnwidth]{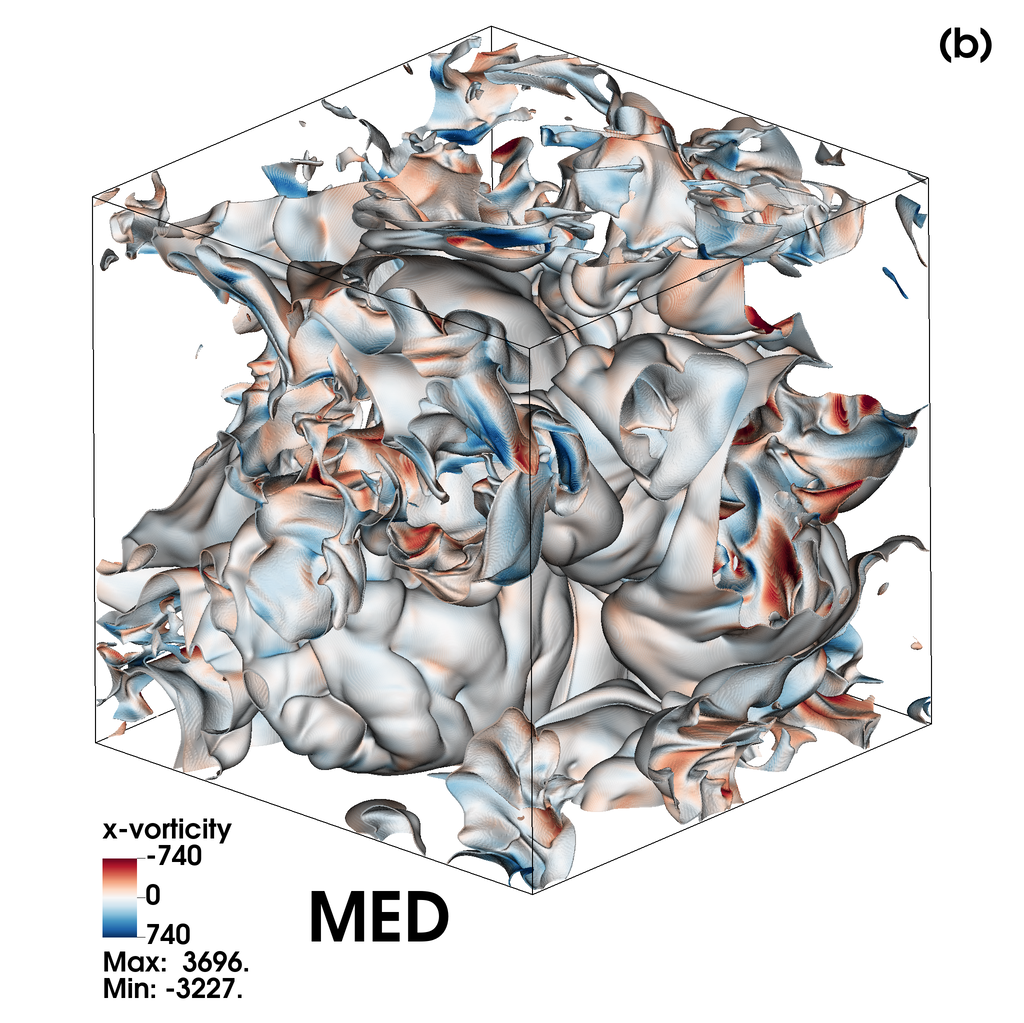} &
        \includegraphics[width=0.64\columnwidth]{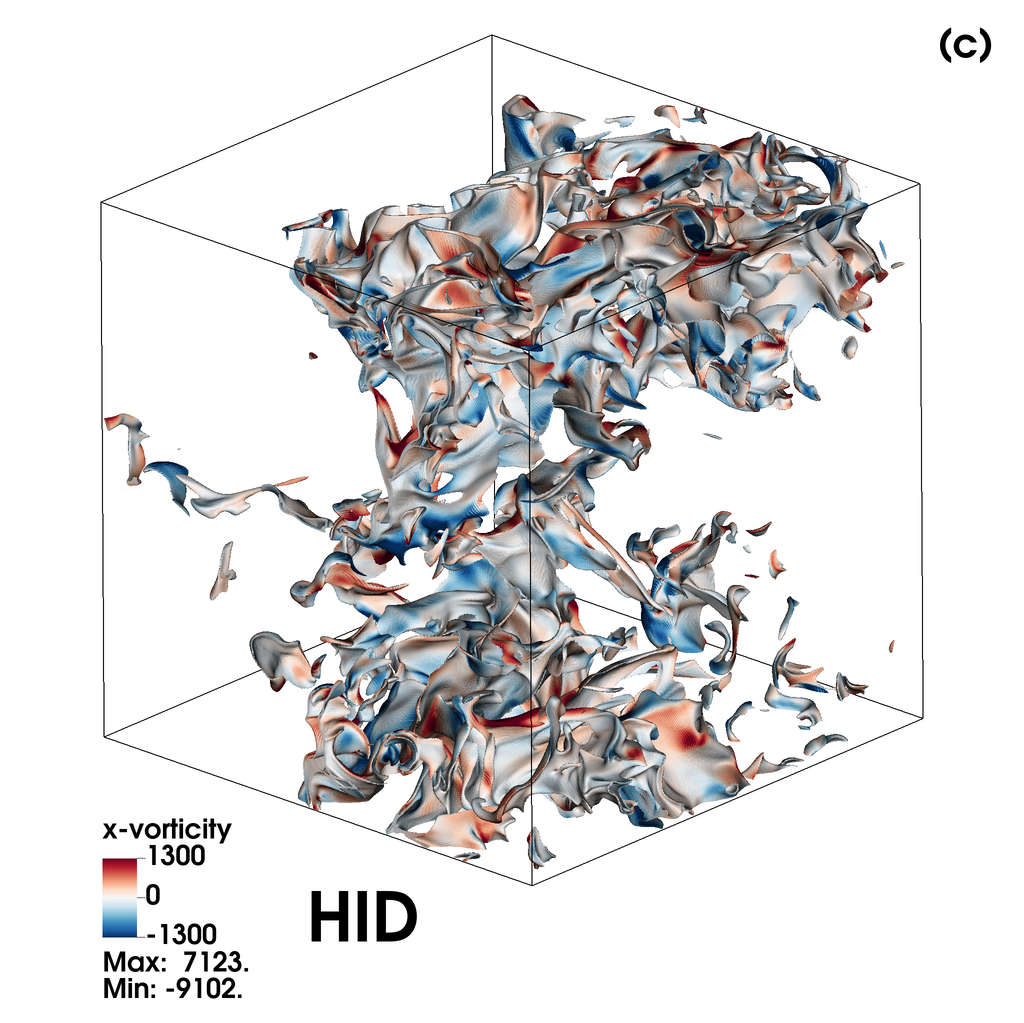} \\
        \includegraphics[width=0.64\columnwidth]{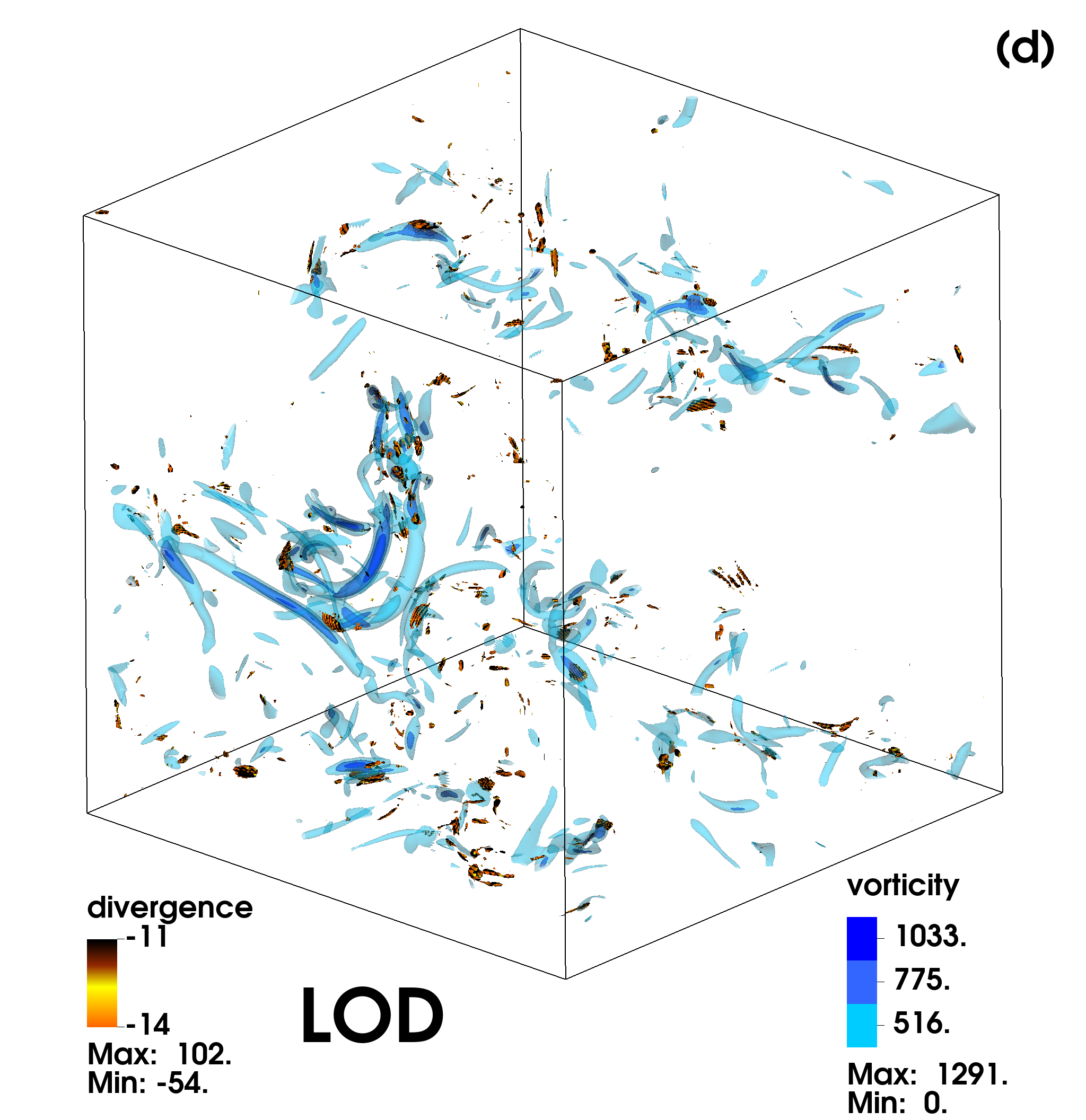} &
        \includegraphics[width=0.64\columnwidth]{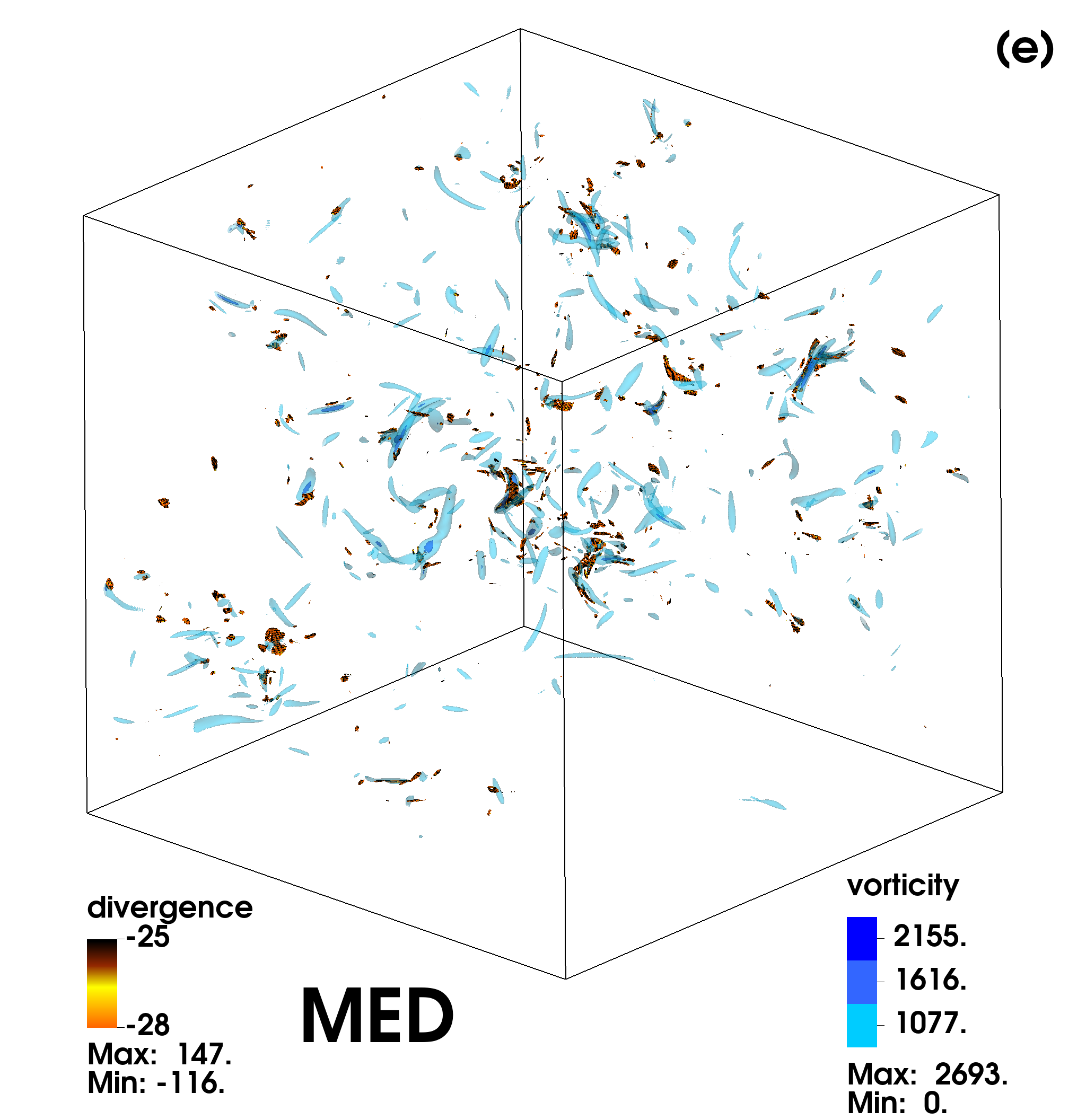} &
        \includegraphics[width=0.64\columnwidth]{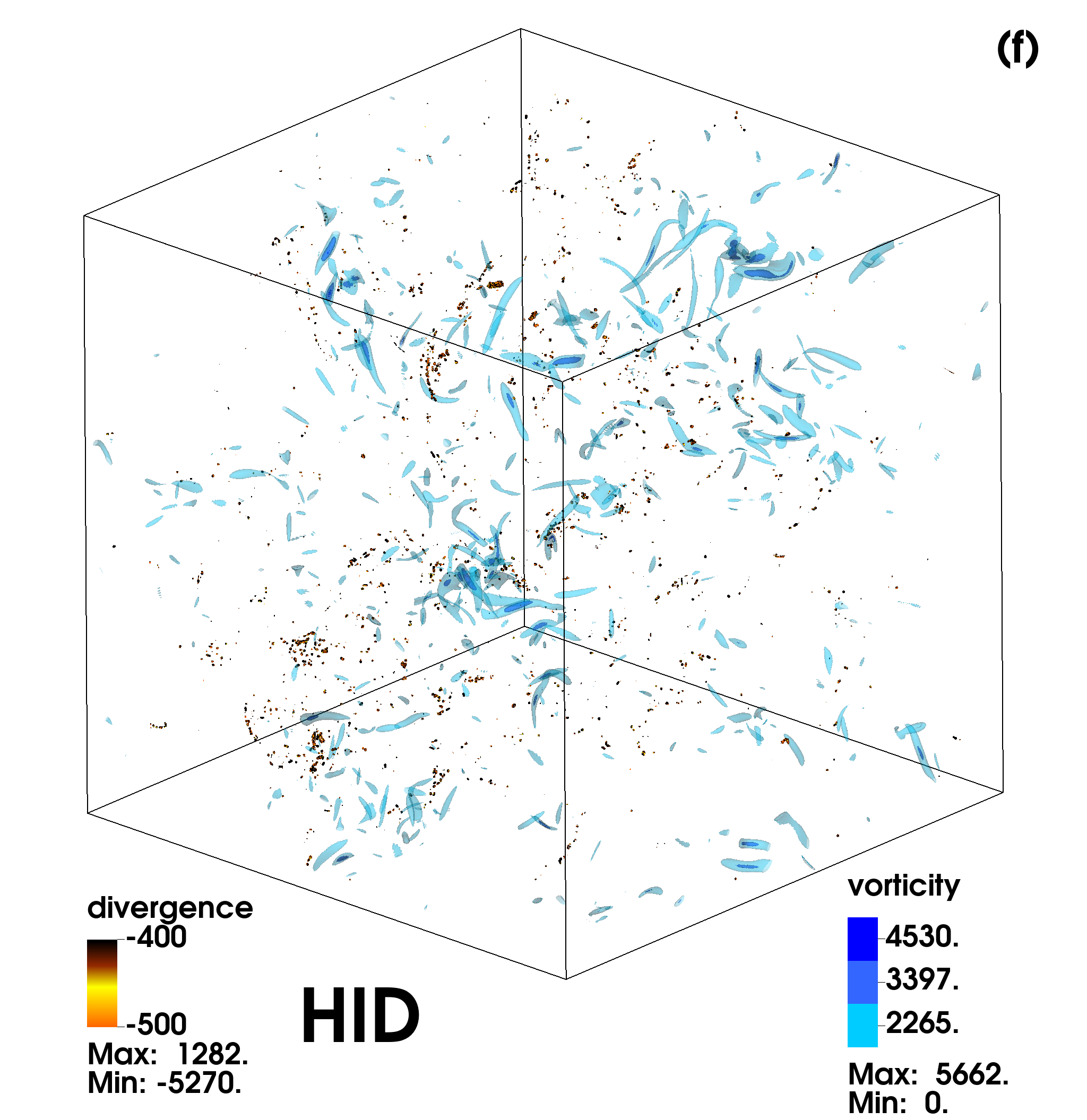}        
    \end{tabular}
    \caption{Model flame morphology and select flow structures inside the TPF region. The low density (LOD), medium density (MED), and high density (HID) model data at the final simulated times are shown in the left, middle, and right columns of panels, respectively. The top row shows an iso-volume of the flame progress variable between 0.4 and 0.6 colored by the x-component of the vorticity, while the bottom row shows the iso-volume of divergence representative of modest compression (varying hues of yellow-red; bright to dark) and contours of vorticity at 40, 60, and 80 per cent of the maximum vorticity (colored with various shades of blue). See text for details.}
    \label{f:morphology}
    \end{center}
\end{figure*}

Fig.\ \ref{f:morphology}
%
%
depicts the flame surface morphology of the turbulent post-flame region (TPF) for each model density. These morphology plots are accompanied by contour plots of vorticity and velocity divergence structures in this region, which shows turbulence becoming somewhat better organized with longer vortex tubes and few but larger shocklets as the density decreases. 

\begin{figure}
    \begin{center}
    \includegraphics[width=\columnwidth]{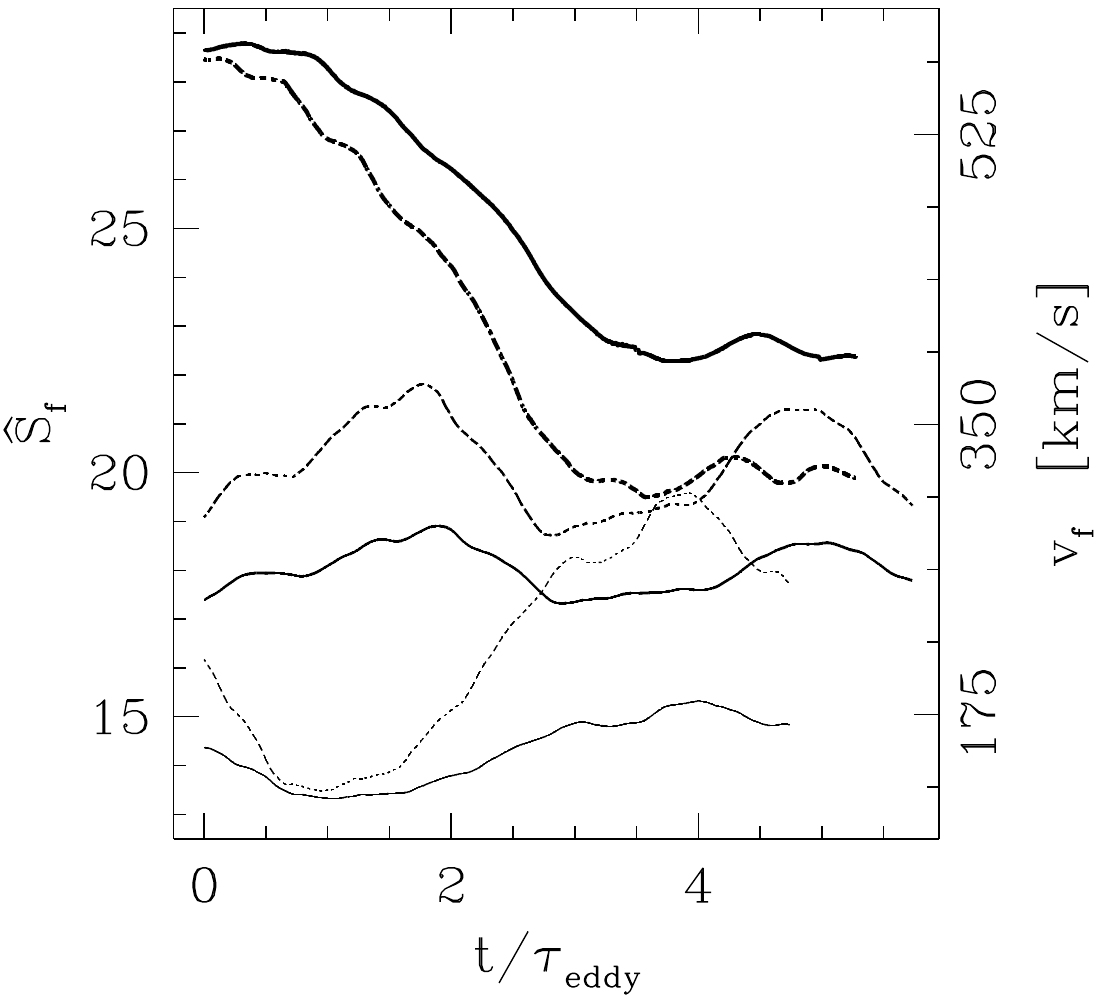}
    \caption{Time evolution of the main turbulent flame properties. Evolution of the flame surface area increase factor, $\hat{S}_\mathrm{f}$, and the turbulent flame speed, $v_\mathrm{f}$, are shown with solid and dashed lines, respectively, for LOD (thin lines), MED (medium lines), and HID (thick lines). The average values are given in Table \ref{t:flameTPF}.
    }
    \label{f:flameSpeed}
    \end{center}
\end{figure}

The flame brush in the TPF region in the LOD model is characterized by rather large bubbles and fairly smooth flame surface with smaller structures intermixed. One can also see the presence of long vortex tubes and small patches of compression. This region in the MED model shows a similarly complex structure, however the flame bubbles appear somewhat smaller and thus less smooth with seemingly richer fine-scale structure. The vortex tubes are also visibly smaller in the MED model. These trends appear to continue as the density increases with the flame surface being progressively more convoluted. In consequence, the flame surface in the HID model is more than 20 times greater than that of a planar flame (see Table \ref{t:flameTPF} and Fig.\ \ref{f:flameSpeed}).
%
%
Also, as the density increases, the vortex tubes decrease in size while, as evidenced by the velocity divergence, acoustic perturbations become stronger. We note that even though the flame surface increase is less in lower density models, it is still very substantial (factor $\approx18$). 

\begin{figure}
    \begin{center}
    \includegraphics[width=\columnwidth]{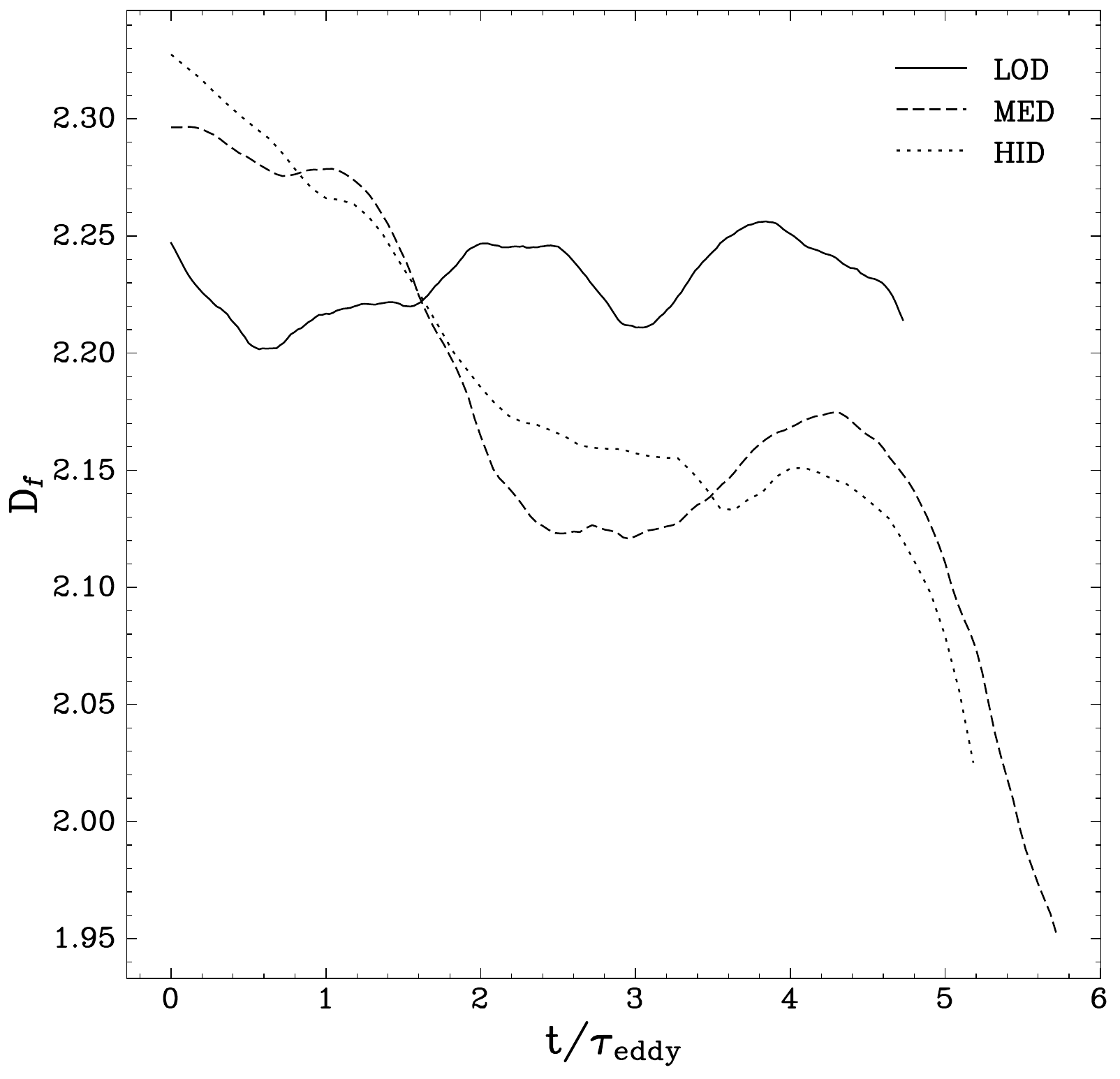}
    \caption{Time evolution of the fractal dimension of the RT-unstable thermonuclear flame models, $D_\mathrm{f}$, in the TPF region. Fractal dimension was obtained for the iso-contour value of the flame progress variable, $\phi=0.5$. The results indicate that the flame surface fractal dimension shows significant variation in time with average value around 2.2 in all models. See text for discussion.}
    \label{f:fracdim}
    \end{center}
\end{figure}

The above-described structure of the flame brush is consistent with the fractal dimension analysis of the flame surface in the TPF region. Fig.\ \ref{f:fracdim}
%
%
depicts the time-evolution of the flame surface fractal dimension in the TPF region of each model. In all model realizations during the analyzed time interval, flame fractal dimension varies by up to 10 percent with the average value of about 2.2. This indicates that the flame structure has predominately a surface-like appearance with some addition of genuinely three-dimensional components, such as vortex sheets. It is also interesting to note that the time evolution of the fractal dimension is reminiscent of the behavior of the fuel fraction during the analyzed time interval (cf.\ Fig.\ \ref{f:fuelfrac}). A likely explanation for the similar behavior of those two quantities could be that the ability of the flame to produce convoluted structures gradually decreases as the amount of flame surface decreases.
\subsection{Overall properties of the gravity-dominated flame turbulence}
\begin{figure}
    \begin{center}
        \includegraphics[width=\columnwidth]{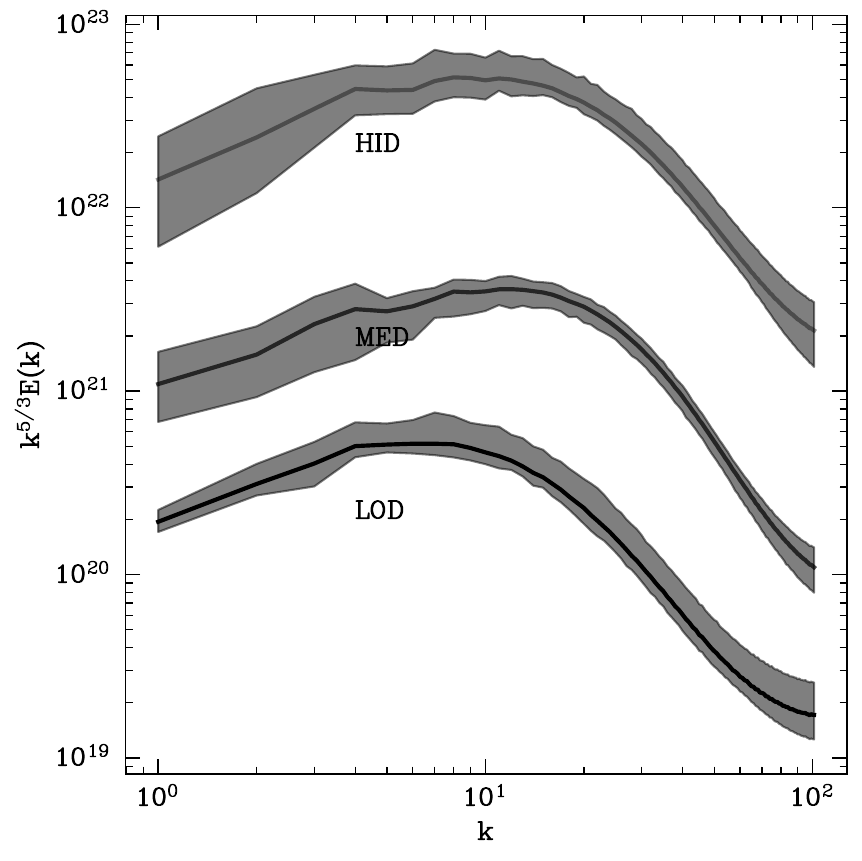}
    \caption{Time-averaged compensated kinetic energy (KE) spectral density inside the TPF region. The KE spectra are compensated by the expected $k^{-5/3}$ Kolmogorov scaling, and show a relatively short, flat region ($k\approx8\pm4$) consistent with the presence of an inertial subrange.}
    \label{f:ke}
    \end{center}
\end{figure}

The source of energy driving turbulence in SN Ia deflagrations is gravitational potential energy of the fuel. Fig.\ \ref{f:ke}
%
%
shows time-averaged compensated kinetic energy spectral density in our models. Our first observation is that the overall shape of the compensated spectra appears consistent with that characterizing well-developed turbulence. Second, the amount of kinetic energy in turbulence varies between the models by about two orders of magnitude. This is due to the difference in terms of the fuel density and the gravity between the models. As a result, the RMS velocity turbulence, as shown in Table \ref{t:turbTPF}, varies from about 200 km/s in the LOD model to almost 700 km/s in the HID model.

Presence of nearly flat segments in the graph of compensated kinetic energy spectral density provides evidence for existence of the inertial subrange, which extends for about one decade starting at $k\approx4$. More formally, the estimated time-averaged integral scale of turbulence in the TPF region, L$_k/$L$_{box}$, ranges from 0.7 to 0.83 as model density decreases. This is consistent with visual appearance of the flame surface showing more fine scale structure, as noted in Section \ref{s:results}. The critical RTI mode and the maximum unstable RTI mode \citep{Takabe1985,LivneArnett1993} were found to vary from about 0.3 to 0.5 and 1.1 to 2.2, respectively, as model density decreases.

The amount of energy available to energize RTI turbulence is given by the amount of buoyancy, which is directly related to the amount of remaining unburned material. It is at its maximum at the leading edge of the flame region and gradually decreases as burning proceeds inside the flame brush. Fig.\ \ref{f:fuelfrac} shows the time evolution of the TPF fuel mass fraction in our models. Given our adopted definition of the TPF region it typically contains about 20 per cent of fuel. However, the fuel amount may significantly vary by up to a factor of four with typical variation of approximately a factor of two. This, coupled with the convoluted morphology of the flame surface, as shown in Fig.\ \ref{f:morphology}, indicates the prevalence of fuel-rich channels embedded within the rising flame bubbles. On the other hand, it also suggests that the drive intensity is likely to substantially vary during the evolution on timescales relevant to large-scale turbulence. Those variations in forcing ultimately do affect the dynamics of the flame brush region. The aforementioned variations of the amount of fuel in the TPF region also seem to be occurring on a timescale comparable to the large-scale turnover time of turbulence, perhaps modified by the ability of the flame brush region to collectively respond to the external forcing.

The time variability of the RTI forcing is illustrated in Fig.\ \ref{f:ke} with filled contours marking the range of variability in the spectral contents of the kinetic energy during the analyzed time period. As we remarked earlier, the buoyant energy is injected at large scales, $k\approx4$, which reflects the presence of RTI bubbles of both large and intermediate size (compared to the lateral extent of the computational domain). It is important to note that the energy is injected with a rate varying up to 50\% in the strongest driven model, HID; the forcing varies about 10 to 20\% in the other two models. Given that the buoyant energy still available in the TPF region amounts to only a few per cent of the initial buoyant energy, we found the turbulent kinetic energy in this region is comparable to, and in some cases may substantially exceed (about 70\% on average in the MED model), the available buoyant energy. This explains the observed highly variable nature of the flame evolution.

\begin{figure}
    \begin{center}
        \includegraphics[width=\columnwidth]{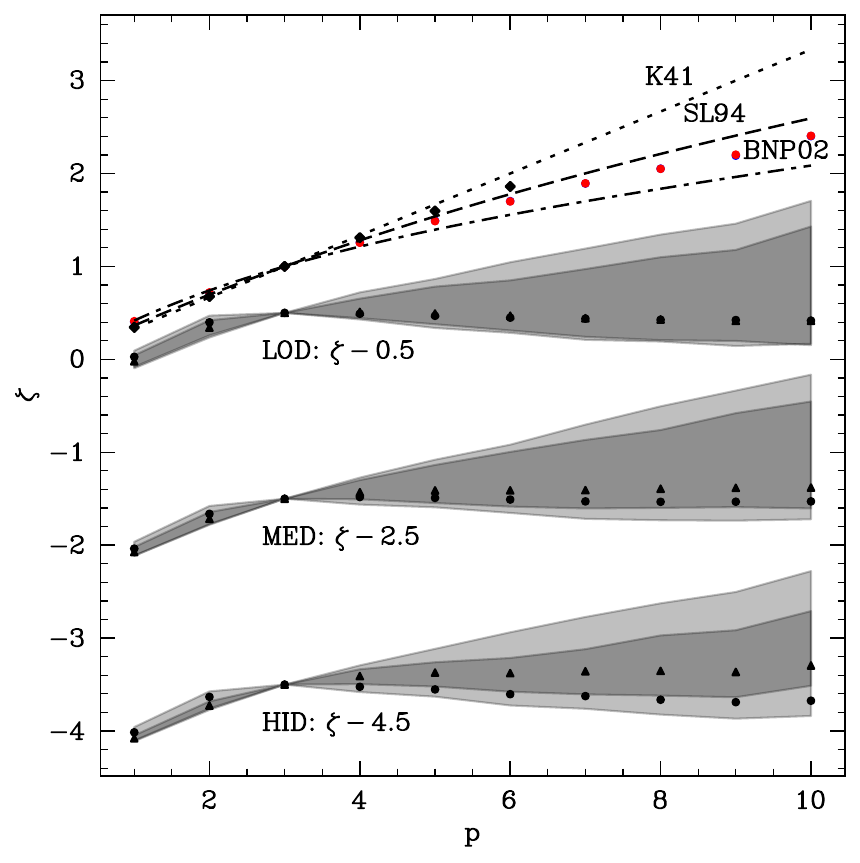}
        \caption{Time-averaged velocity structure functions (VSFs) inside the TPF region. The individual longitudinal and transverse VSFs are normalized by their respective third-order VSF. The gray-shaded bands highlight the variations between longitudinal (traced by disks) and transverse VSFs (marked with triangles). The VSF scalings appear saturated for orders greater than three, and strongly deviate from theoretically predicted scaling laws by \citet[][K41]{Kolmogorov1941}, \citet[][SL94]{She1994}, and \citet[][BNP02]{Boldyrev2002}. Representative values for VSF scaling exponents found by \citet[][Table 1]{CiaraldiSchoolmann2009} in SN Ia explosion model of \citet{Roepke07} are shown with diamonds. The exponents reported by \citet{FennPlewa2017} for SD models are marked with red disks. For clarity of presentation the LOD, MED, and HID model flame VSF results are offset by $-0.5$, $-2.5$, and $-4.5$, respectively. See text for discussion.}
    \label{f:vsf}
    \end{center}
\end{figure}

It is conceivable that the observed variability of RTI-turbulent drive may affect turbulence beyond the spectral distribution of kinetic energy. As far as the structure of turbulence is concerned, we first consider two-point correlation statistics expressed through velocity structure functions (VSF). Fig.\ \ref{f:vsf}
%
%
shows the dependence of the scaling exponents as a function of VSF order in our series of models. In order to facilitate comparison with existing literature, we normalized the exponent values of both longitudinal and transverse VSFs to the 3rd-order VSF. These are shown with filled circles and filled triangles, respectively. Their time variability is indicated by semi-transparent filled contours colored with light gray. Finally, for clarity of presentation, the VSF model data was shifted by about $-0.5$, $-2.5$, and $-4.5$ for LOD, MED, and HID models, respectively. For comparison, we are also providing VSF scaling relations theoretically predicted by \citet[][short-dotted line; K41]{Kolmogorov1941}, \citet[][long-dashed line; SL94]{She1994}, and \citet[][dash-dotted line; BNP02]{Boldyrev2002}. 

We observe a striking, qualitative difference between the RTI flame VSF scalings and broadly considered theoretical predictions for isotropic turbulence. The time-averaged RTI VSF scaling exponents saturate for $p>3$ and do not appreciably change their values as the order increases. There is also noticeable difference between the longitudinal and transverse VSF scaling exponents, particularly in the two higher density models. We are not in a position to evaluate statistical significance of these differences as we did not consider time-dependent variations of these exponents in required detail. We speculate that those differences might be a consequence of gravity playing a gradually weaker role as the density decreases.

The longitudinal VSF scaling exponents in all flame models appear to be gradually decreasing with the order. This particular trend can be attributed to a highly variable driving mechanism and, in consequence, strongly intermittent nature of turbulence realized in this problem. Also, compared to our results obtained for homogeneous, isotropic turbulence \citep[BFP21,][]{FennPlewa2017}, the difference between longitudinal and transverse scalings might be due to gravity-induced anisotropy of the drive. For example, a simple measure of anisotropy of turbulent velocities in the TPF,
\[
\chi_\mathrm{v} = \frac{v_\mathrm{RMS,x}}{\sqrt{v_\mathrm{RMS,y}^2 + v_\mathrm{RMS,z}^2}},
\]
as shown in Table \ref{t:turbTPF}, is close to 1.2 in all models, and similar to the anisotropy reported by \citet[][see his Fig.\ 10]{Khokhlov1995}, but exceeds factor of 2 for much denser fuel. This points to an overall anisotropic turbulent motions in the TPF with fuel-rich RT spikes retaining some of their initial ballistic dynamics while they sink through the flame brush. Similar anisotropy of turbulence dynamics is reflected in the radial component of the Reynolds stress tensor being larger than the lateral tensor components (cf.\ Section \ref{s:tpf}).

The extreme ranges of the scaling exponents, as marked by gray color-filled contours in Fig.\ \ref{f:vsf}, indicate that the TPF VSF scalings remain in the saturated state most of the time (closeness of the lower extreme range of the exponent values to their time-averages). Also, the scaling of the exponents appears bounded by both theory \citep{She1994,Boldyrev2002} and predictions of relevant computer models as the slope of the upper extreme range envelope approximately agrees with our previous findings \citep{FennPlewa2017}. Clearly, the observed behavior of the VSF scalings is anomalous and strongly differs from the behavior originally predicted by \citet{Kolmogorov1941} and the extended similarity models such as \citet{She1994}. As elucidated by \cite{Yakhot1998} following the ideas of \cite{Polyakov1993} and \cite{Landau1987}, the observed saturation of the scaling exponents can be due to the time-variability of the drive.

\begin{figure}
    \begin{center}
    \includegraphics[width=\linewidth]{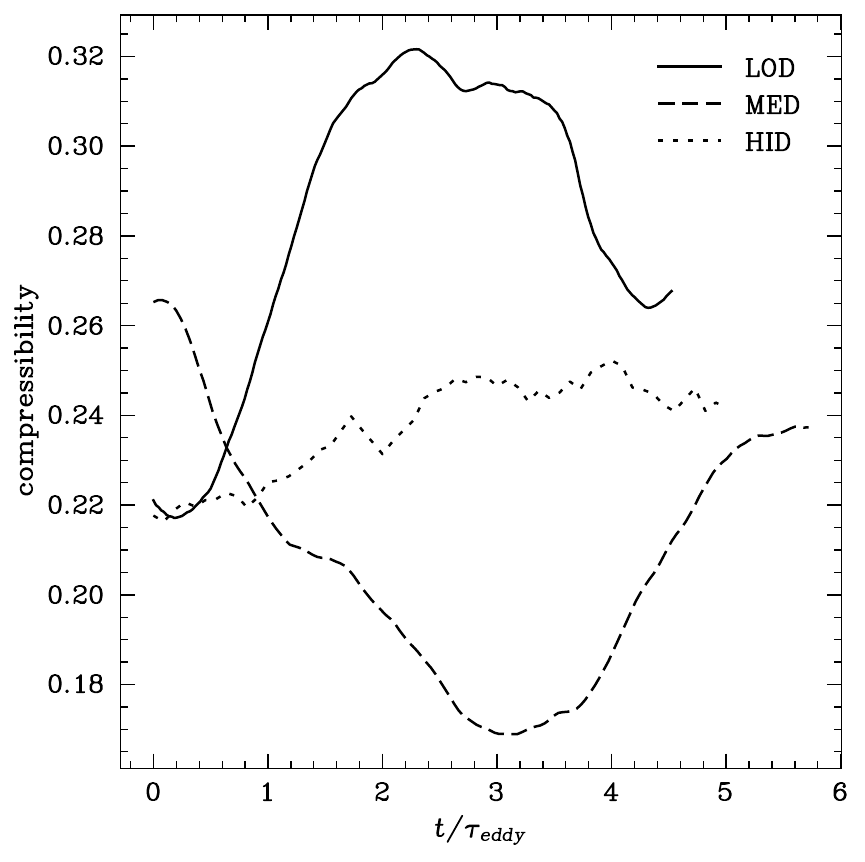}
    \caption{Evolution of the compressibility of TPF turbulence as a function of time shown in units of eddy turnover time, $\eddyt$. The amount of compressibility in the LOD, MED, and HID models is depicted with solid, dashed, and dotted lines, respectively. See text for discussion.}
    \label{f:compressibility}
    \end{center}
\end{figure}

The aforementioned drive variability may also impact other statistical turbulence measures. As recently demonstrated by \citet[][see also \citet{Mouri2006}]{Bentkamp2025} in their study of three-dimensional isotropic and homogeneous turbulence, time-dependent drive increases high-order moments, such as flatness and hyper-flatness, in particular for velocity. Using the Helmholtz decomposition, we found a substantial amount of compressibility in the TPF turbulent velocity field. This is illustrated in Fig.\ \ref{f:compressibility},
%
%
which shows that typically compressive modes represent about 25\% of the TPF velocity field, with the extreme compressibility ranging between about 20\% and 30\%. This relatively large amount of compression exists despite the fact that the RTI drive itself is essentially incompressible -- the assumption routinely made when studying evolution turbulent deflagrations on small-scales \citep[see, for example,][and references therein]{Aspden2010}. The amount of the TPF compressibility appears varying on relatively long timescales that might potentially be related to the quasi-periodic nature of the RTI-drive employed in our simulation setup, as indicated by the time evolution of the flame speed (cf.\ Fig.\ \ref{f:flameSpeed}). Presence of those longer timescales in our models is consistent with the findings of \citet{Zhang2007}, who used a similar flame-in-a-box setup with initial conditions closely matching our HID model and reported a typical timescale of 200 ms for flame evolution on large scales.
\subsection{Fuel properties in the TPF region}
So far our focus was on characterizing RTI-driven turbulence operating in the post-flame region, and comparing its properties to that of our earlier SD turbulence models. This comparison was motivated by the expectation that if both systems produce on average turbulence of similar characteristics, then one would naturally expect to observe similar outcomes. In the context of the ZDDT mechanism, the SD models showed that a suitable preconditioning of fuel is possible. We attributed this process to an adiabatic compression of fuel in channels bounded by flame segments by meso-scale turbulence. The question then is whether a similar behavior is present in our RT-unstable flame models.

\begin{figure*}
    \begin{center}
    \begin{tabular}{ccc}
        \includegraphics[width=0.66\columnwidth]{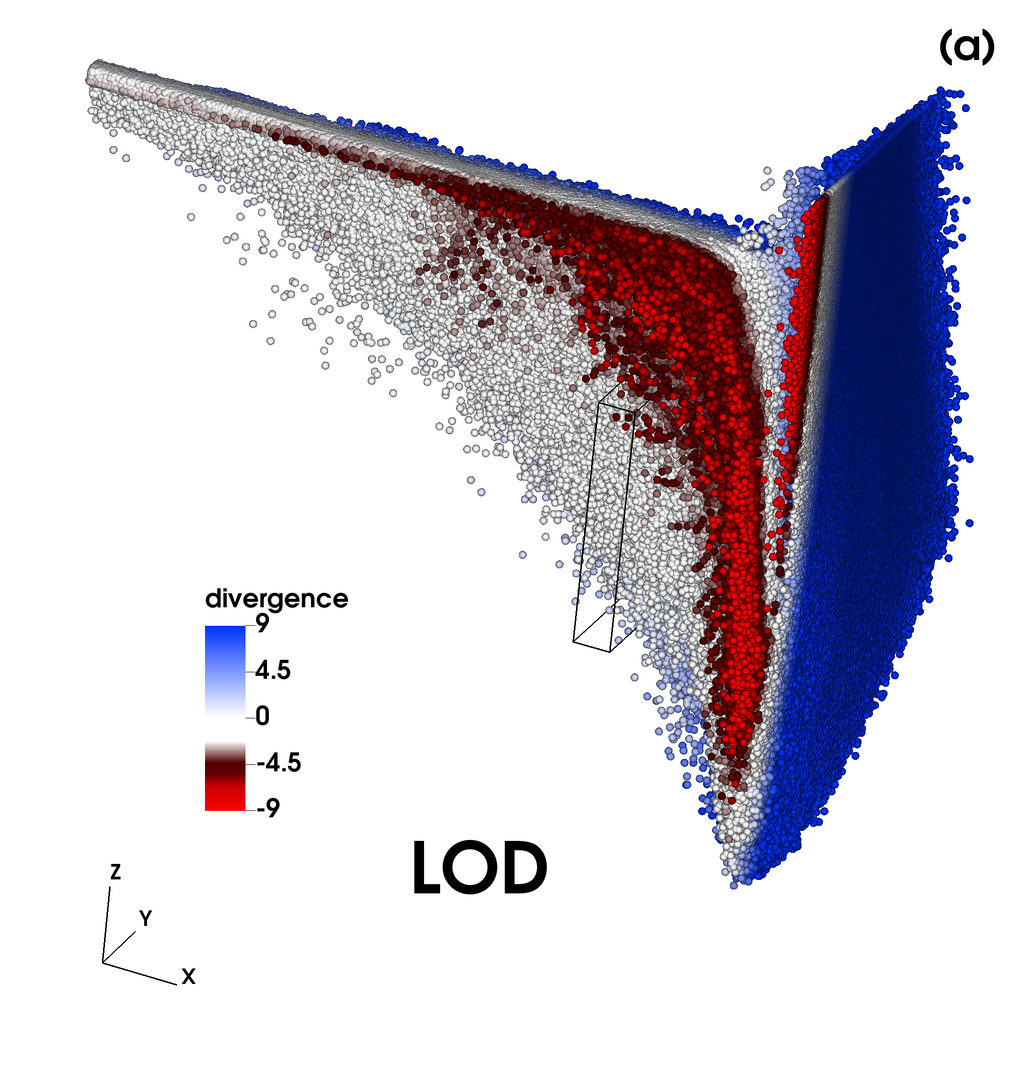} &
        \includegraphics[width=0.66\columnwidth]{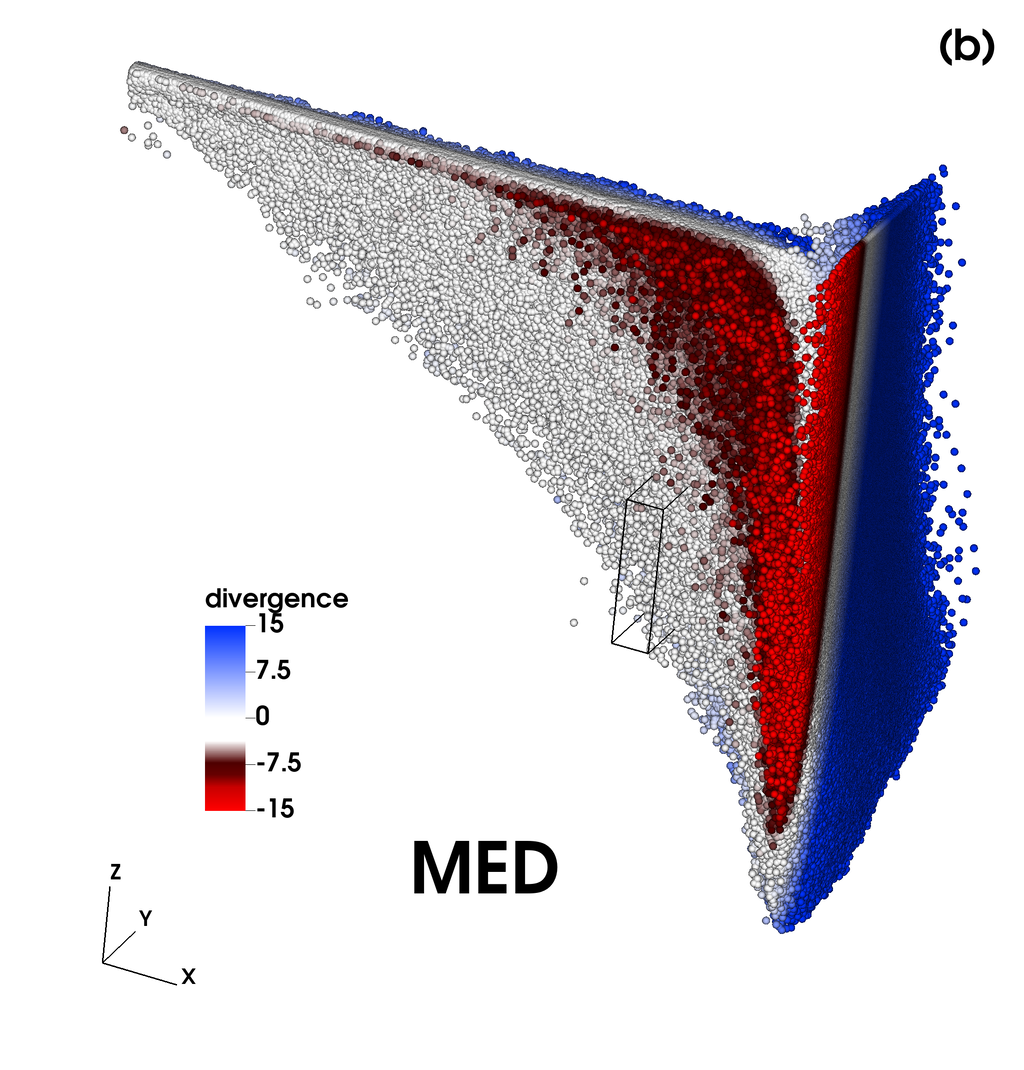} &
        \includegraphics[width=0.66\columnwidth]{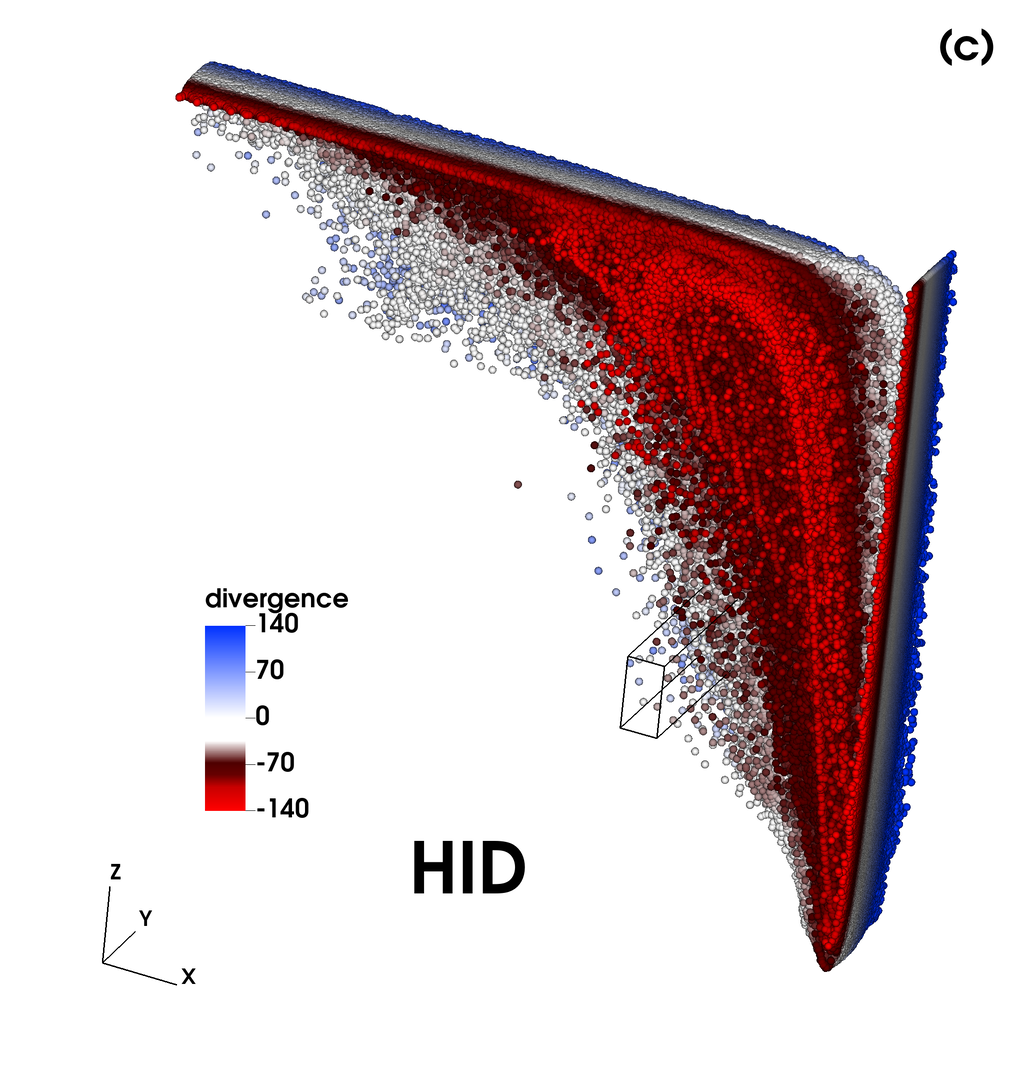}
    \end{tabular}
    \caption{Scatter plot rendering of the individual mesh cells in the TPF region in (a) LOD, (b) MED, and (c) HID flame models at a single time slice. The mesh cell distribution is shown in terms of flame progress variable in $\log_{10}$-scale along x-axis, velocity divergence along y-axis, and (carbon) ignition time in $\log_{10}$-scale along z-axis. (The axes in these plots should not be confused with coordinate axes in the hydrodynamic simulations.) The log-scaled flame progress variable ranges from $-20$ (pure fuel) to 1 (complete burn) in all panels, while the data in the other two dimensions are individually scaled to form a cube of equal dimensions. Note that the individual points are colored by velocity divergence with red indicating compression and blue indicating expansion. The color range varies between panels and was adjusted to highlight a population of cells undergoing compression as the flame front advances toward them. See text for details.}
    \label{f:scatter}
    \end{center}
\end{figure*}

Fig.\ \ref{f:scatter} 
%
illustrates the state of TPF plasma in terms of the log$_{10}$-scale of the flame progress variable (increasing along the x-axis; the amount of carbon fuel is proportionally decreasing in that direction), the velocity divergence (centered around zero and increasing along the y-axis), and the (carbon) ignition time (in terms of log$_{10}$-scale; increasing along the z-axis). In each figure panel, a wire-frame slab is embedded in scatter plot with its x-segment covering the region of the flame progress variable, $\log_{10}\phi=[-5,-4]$, and extending by one order of magnitude in terms of the ignition time (in the z-direction). We note that in all models considered, the distributions have a similar form, which we interpret as follows.

Because due to numerical diffusion our scheme produces a diffusive precursor, there exists a relation between the flame progress variable and the physical distance across the flame front such that small values of the flame progress variable indicate large distances from the thermodynamically active central segment of the flame profile. In Fig.\ \ref{f:scatter}, material well separated from the flame front occupies top-left segments of the distributions located about 1 km (16 mesh cells) away from the region where most of the burning energy is deposited.\footnote{At relatively low densities of the stellar plasma considered in this work, almost all energy delivered by the 3-stage reactive flame module of the ADR flame model is deposited when the flame progress variable, $\phi$, ranges between 0.1 and 0.9; see discussion in Section \ref{s:methods} and \citet{Khokhlov1991a}.} At these large distances from the front, the essentially pure fuel remains unaffected by the flame dynamics with the velocity divergence close to zero. At the same time, the ignition time does not appear significantly affected either, remaining close to that of the unperturbed, cold fuel.

\begin{figure*}
    \begin{center}
    \includegraphics[width=\linewidth]{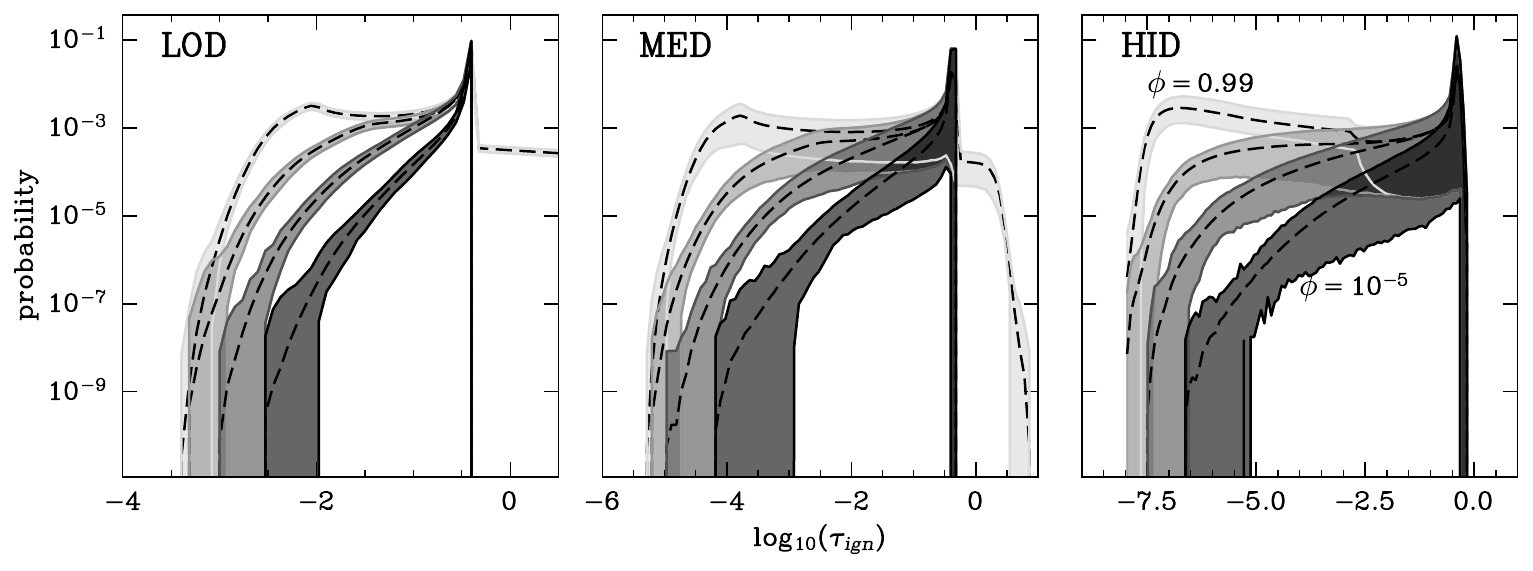}
    \caption{Time-averaged PDFs of ignition time for the material in the TPF region constrained in terms of the flame progress variable for the LOD (left panel), MED (middle panel), and HID (right panel) model. Each time-averaged PDF is shown as a dashed line, and the associated gray-shaded band depicts the extreme range of the ignition time within the analyzed time-series. The threshold values defining the maximum value of the flame progress variable used to construct the PDF are indicated in the right panel for two extreme cases. The threshold values used are $1\times10^{-5}$, 0.01, 0.1, 0.99, colored from lightest gray to darkest gray. See text for discussion.}
    \label{f:pdfs1d}
    \end{center}
\end{figure*}

With time, the aforementioned quiescent state of the fuel gradually changes as the flame front approaches the fuel parcels, and both dynamical effects and thermodynamical state become more diverse. In particular, the fuel ignition time decreases by about two orders of magnitude in the LOD model (as indicated by a light gray-filled contour in the leftmost panel of Fig.\ \ref{f:pdfs1d}),
%
%
by about three orders of magnitude in the MED model (same-colored contour in the middle panel of the same figure), and by more than five orders of magnitude on average in the HID model (same-colored contour in the rightmost panel of the same figure).\footnote{Table \ref{t:modelICs} provides ranges of ignition times recorded in our models. The initial ignition time in all models was $\approx330$ ms.} We attribute this change in the ignition time to the physical proximity of the fuel to the active flame front when the flame thermal expansion results in the adiabatic fuel compression. In Fig.\ \ref{f:scatter}, the material that undergoes compression is colored in various shades of red. This population of cells arches down indicating a progressive decrease in the ignition time as the distance to the flame front decreases.

Besides the overall gradual evolution of plasma due to approaching flame front discussed above, we also identify presence of fuel  which is more strongly than on average impacted by the flame front. This material is indicated by isolated, nearly linearly aligned groups of red-colored points in Fig.\ \ref{f:scatter}. These groups are especially well visible in the HID model (rightmost panel in the figure) and to a lesser extent in lower density models. We propose that the presence of those distinct groups is a manifestation of particularly intense interactions between the flame front segments with the fuel. That those groups appear as elongated, seemingly connected chains of data points in Fig.\ \ref{f:scatter} indicates the presence of coherent flame interaction events, which are correlated in time and space. They collectively affect the fuel dynamics (due to similarity of compression) and its thermodynamics (due to similarity in terms of the ignition times). In other words, small segments of the flame front create populations of plasma samples that display highly correlated behavior, seemingly in bursts of activity that strongly impact the fuel.

Time-averaged ignition time PDFs of the fuel provide additional evidence for the fuel-flame interaction. Assuming at first that no such interaction takes place, the fuel should be entering the flame front at its upstream thermodynamic state closely matching the initial conditions. Clearly, this is not the case in the RT-driven turbulent flames, as evidenced through a strongly asymmetric form of fuel ignition time PDFs shown in Fig.\ \ref{f:pdfs1d}. Here we consider fuel ignition time PDFs conditioned on the value of the flame progress variable, which, given the aforementioned diffusive flame precursor, effectively corresponds to various distances away from the flame front. In all our models, the material with the longest ignition time, which one could consider as the cold fuel phase, constitutes the most populous set with PDF reaching a level of a few per cent.\footnote{We note that in LOD and MED models the presence of two populations of plasma samples representing the final stage of carbon burning ($\phi=0.99$) with the longest ignition times (depicted in the figure by tails extending towards long ignition times with probability $\approx$\num{1e-4}), are numerical artifacts of the formula used to calculate carbon ignition times \citep{DursiTimmes2006}. This formula produces large values of ignition times where carbon is nearly exhausted.} As the flame approaches, a population of plasma parcels (marked with a dark gray filled contour in the figure; $\phi=1\times10^{-5}$) emerges from the cold fuel phase and forms an extended tail with ignition times reduced from about two to five orders of magnitude in LOD and HID models, respectively, with an intermediate degree of ignition time shortening in the MED model. Those tails having similar shapes is pointing toward a common physics mechanism being responsible for the shortening of fuel ignition times well ahead of the flame front.

The aforementioned flame-fuel interactions appear highly dynamic in nature, as evidenced by significant broadening of the time-averaged fuel ignition time distribution tails. The same mechanism seems to continue to operate as the flame front approaches with a gradual continuing decrease in fuel ignition times (medium-dark and medium-light gray filled contours in Fig.\ \ref{f:pdfs1d}). As the fuel enters the flame front proper (medium-light and light gray filled contours in the figure), we no longer observe significant shortening of ignition times, with a nearly completely burned material indicated by a distinct secondary maximum in the region of very short ignition times.
\section{Discussion}\label{s:discussion}
In the remaining part of this paper, we first focus on comparing conditions that exist in the RTI-driven turbulent flame to those considered in spectrally-driven (SD) models of turbulent thermonuclear combustion reported by BFP21, which produced DDT via the Zel'dovich mechanism. We then evaluate our findings in the context of previous studies of turbulent flames in Type Ia SNe, including RT-driven flames \citep{Bell2004, Zingale+05, Woosley+11} and flame-turbulence interaction \citep{Poludnenko2019}.
\subsection{Fuel heating and preconditioning}
BFP21 found that preconditioning required for DDT was produced inside relatively narrow channels of fuel bounded by deflagration fronts. Initial conditions and numerical setup adopted in this study aim to closely match the parameters of SD models so that the main difference between the two studies, and thus a potential culprit for differing outcomes between them, could be associated with different turbulence driving mechanisms.

Our preference for comparing the two models in terms of turbulence properties, and thus necessarily limiting our interest in the RTI model to a restricted post-flame region, does not preclude the possibility of the DDT preconditioning process taking place in other parts of the post-flame region. Although such a possibility cannot be completely ruled out, fuel preconditioning requires a high level of perturbations that naturally develop as RTI energizes the flow, so that post-flame region is the most likely locale for DDT.

\begin{figure*}
    \begin{center}
    \includegraphics[width=\linewidth]{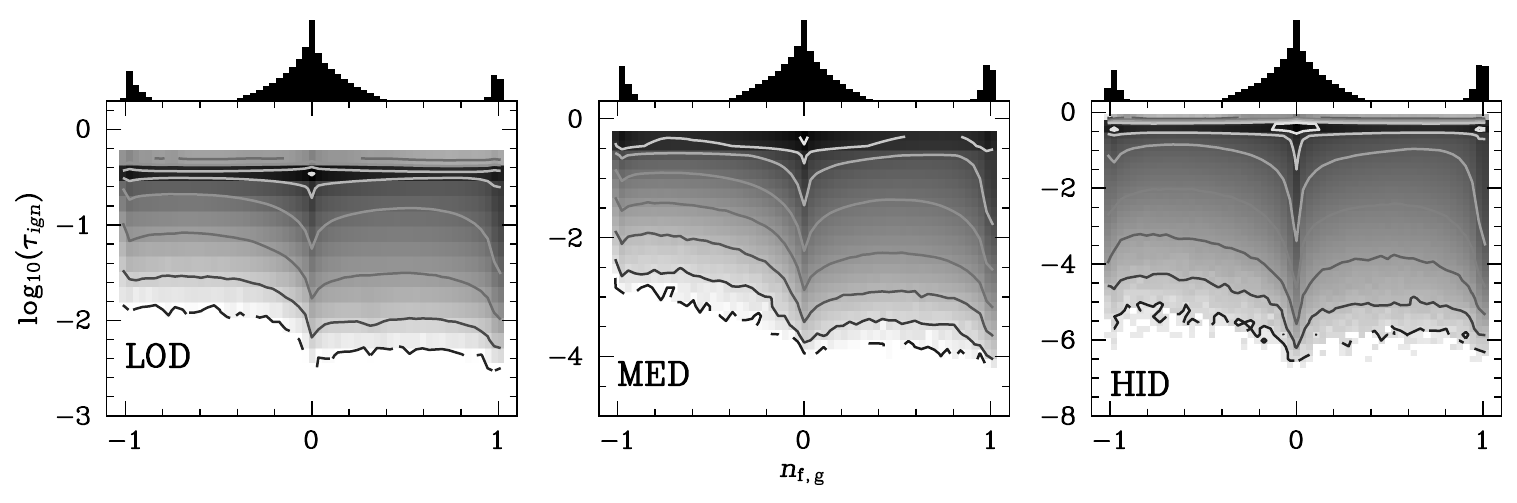}
    \caption{Pseudo-color map of time-averaged joint PDFs of ignition time for the fuel layer adjacent to the flame front, shown in $\log_{10}$-scale, and the normalized gravitational acceleration projected in the direction normal to the flame surface, $n_\mathrm{f,g}$. The projected acceleration is maximal at the tops of rising flame bubbles. The PDFs were obtained for the unburned material corresponding to the flame progress variable, $\phi\leq\num{1e-5}$. The highest probability values are indicated by dark-gray flooded contours while the lowest probability values are indicated by light-gray flooded contours. The histograms attached to the top of each panel show the relative populations of analyzed fuel samples. Note that the ignition time ranges change between panels, and ignition times for fuel samples associated with flame surface segments accelerated by buoyancy force, corresponding to data with $n_\mathrm{f,g} > 0$, are systematically shorter than in other regions of the flame front, perhaps except the flame cusps ($n_\mathrm{f,g}\approx=0$). See text for discussion.}
    \label{f:dphi2dpdf}
    \end{center}
\end{figure*}

The overall evolution of fuel ignition times as the flame front approaches could be understood as follows. First, consider a plasma parcel of pure fuel. As the flame front approaches, the value of the flame progress variable for that plasma parcel gradually increases. In the case of a flat flame front surface, any dynamic effects due to flame front approach would affect all fuel parcels in the same way. In consequence, that population of plasma parcels would be characterized by a single ignition time value. However, in the case of the deformed flame surface, the dynamic impact of the buoyancy force is highly non-uniform along the front. This is illustrated in Fig.\ \ref{f:dphi2dpdf},
%
%
which shows that the ignition times inside the fuel layers adjacent to the flame surface are the shortest when flame surface orientation relative to the gravitational acceleration, expressed here as a suitably normalized factor,
\[
n_\mathrm{f,g} = \frac{\nabla\phi\cdot\mathbf{g}}{\norm{\nabla\phi}\norm{\mathbf{g}}},
\]
is positive and close to its maximum value of +1. This extreme value is attained at the tops of rising flame bubbles when the two vectors point in the same direction. In other words, we observe a systematic shortening of fuel ignition times when fuel is compressed by the buoyancy force generated near the tops of flame bubbles. As bubbles overturn, the buoyant acceleration gradually decreases. In the other extreme, fuel in RTI spikes penetrating into the flame brush, characterized by negative values of the surface orientation factor, shows comparatively longer ignition times.

A process of bubble overturning and flame folding will result in deceleration and adiabatic cooling, and increased ignition times of fuel elements that previously were heated by buoyancy. This effect produces asymmetry between left and right segments in panels in Fig.\ \ref{f:dphi2dpdf}. However, due to the problem geometry, every bubble overturn and flame folding event will result in another flame segment becoming RTI-unstable. In consequence, ignition times of fuel elements adjacent to the flame front will on average decrease with time, as seen in Fig.\ \ref{f:dphi2dpdf}. Our conclusions are supported by a large number of fuel samples. This is evidenced by histograms attached to the tops of the figure panels showing the relative number of fuel samples inside the layers adjacent to the flame front in terms of the orientation factor, with both extremes being well sampled.

We observe additional shortening of ignition times for fuel associated with the flame surface segments nearly aligned with the direction of gravitational acceleration. It is conceivable that these fuel samples reside inside the flame cusps that form when flame structures (such as RTI bubbles) collide.\footnote{Evolution of the cusp regions, in particular because of their elongated geometry and potential role in ZDDT, deserves a separate study. For a qualitative discussion of their role in the flame evolution see \cite{Khokhlov1993} and \cite{Hicks19}.} These regions most closely correspond to fuel channels found in our SD models, in which fuel was preconditioned by mesoscale compressive turbulence modes and produced ZDDT.

In addition to systematic shortening of the ignition times of the fuel layers adjacent to the flame front, the cumulative action of the buoyancy force also results in substantial broadening of an initially narrow distribution of ignition times characterizing cold fuel. This is clearly illustrated in Figs.\ \ref{f:scatter} and \ref{f:pdfs1d} with a distribution tail greatly extended toward shorter ignition times. In our models, that tail has a width of about two orders of magnitude in the LOD model and increases to about five orders of magnitude in the HID model in terms of ignition times.

If the characteristics of the fuel layers with the shortest ignition times are similar to the fuel preconditioned by turbulence in SD models, the ZDDT process described in BFP21 might still be viable. One important difference between the RTI and SD models is that fuel parcels adiabatically heated by buoyancy may not have enough time to ignite before being consumed by the flame front. Also, the DDT problem geometry is different with RTI producing a layer of warm fuel rather than an extended section of a fuel channel found in SD models. 

In the case of current models, and as mentioned earlier, some of the strongest dynamic interactions between the flame and fuel occur in regions where the flame fronts collide, forming cusps filled with adiabatically heated plasma. Again, following possible ignition of fuel in cusp regions or layers adjacent to the flame front, the ZDDT mechanism requires a sufficiently long time for a newly born deflagration to develop into a detonation. However, addressing this intriguing problem of ZDDT with preconditioning due to flame-fuel interaction requires detailed direct numerical simulations, and is beyond the scope of the present work.

There are some additional possible consequences of adiabatic fuel heating. It may affect the rate of heat transport across the leading edge of the flame front, modify its thermal structure, and possibly broaden the flame front. In situations when there is sufficient time for a preheated material to ignite, a population of isolated deflagrating regions will be created at some distance from the flame front proper. As those regions ignite and the degeneracy is lifted, each region will produce a spherically expanding acoustic perturbation. Those perturbations may, in principle, additionally enhance burning in the warm fuel layer ahead of the flame front further increasing possibility of fuel ignition.  We observed such positive feedback in SD models (BFP21), and a gradual strengthening of acoustic perturbations toward detonation is one component of the tDDT model \citep{Poludnenko2019}. As in the case of potential formation of ZDDT mentioned in the previous paragraph, studies of additional effects caused by adiabatic fuel heating will also require performing detailed numerical simulations.
\subsection{Comparison to previous SN Ia RTI-driven turbulent flame models}
Many of the investigations into the SN Ia explosion mechanism were based on the results of numerical simulations that by now are 30 years old \citep{Khokhlov1995, Khokhlov1997b}. In later studies, \cite{Lisewski2000}, \cite{Woosley+07}, and \cite{Woosley+09} considered a possibility of triggering detonations by turbulence in the FSF regime with help of empirical 1D turbulence models. Persistence of uncertainty, in particular in terms of inherently intermittent character of turbulence \citep{Pan+08} and limitations of available relevant multidimensional numerical models \cite{Roepke07}, prevented researchers from making definite statements regarding viability of DDT in SN Ia, and motivated research in new directions \citep{Poludnenko2019}. Although results produced by those newly introduced models are encouraging, they were obtained using a set of assumptions that require greater scrutiny.

In what follows, we contrast our results with the previous work done on RTI-driven turbulent flames and recent work on flame-turbulence interaction.
\subsubsection{Zel'dovich-type ignition in RTI-driven turbulent flames}
As discussed in Section \ref{disc_genflamchar}, overall flame evolution in our models during the quasi-steady state phase closely resembles that reported originally by \cite{Khokhlov1995} and more recently by \cite{Zhang2007}. This is not surprising because the adopted initial conditions and simulation setups used by these authors differed only marginally from those used in this study. For example, compared to \cite{Khokhlov1995}, who considered the same initial density as in our HID model, the Froude number in our model is $\mathrm{Fr}\approx 1.14\times10^{-4}$ versus $\mathrm{Fr}\approx 1.88\times10^{-4}$ in the Khokhlov's A model, when adjusted to the same domain lateral size. In consequence, our work, along with that of \cite{Khokhlov1995} and \cite{Zhang2007}, all consider turbulence with strong buoyancy effects \citep{Verma2018}. The two earlier studies indicated a tendency of the RTI driving turbulence on the scales comparable to the size of the domain with essentially only a single RT bubble laterally filling the computational domain. The same tendency is seen in our current models. Therefore, all considered flame-in-a-box models appear to produce very similar results.

The original work by \cite{Khokhlov1995}, along with the first detailed one-dimensional DDT study by \cite{Khokhlov1997b}, provided motivation for \cite{Lisewski2000}, who considered the role of turbulence in the evolution of carbon-rich deflagrations. The main conclusion from their investigation was that much higher intensity of turbulence, on the order of 1000 km/s on a scale of 10 km, than observed in the first generation of multidimensional explosion models, 100 km/s at 10 km as quoted by \citet{Khokhlov1997b}, was required to support a ZDDT scenario envisioned by Khokhlov and collaborators. Several years later, \cite{Roepke07} obtained the next generation of centrally-ignited SN Ia deflagration model, which offered substantially improved resolution compared to the \cite{Khokhlov1995} model, and estimated the maximum turbulence intensity on the order of 1000 km/s at 10 km scale. He concluded that the conditions chosen in \cite{Lisewski2000} may be achieved in SN Ia, albeit perhaps at most a few times per explosion event. The later study by \cite{Pan+08} highlighted the role of intermittency in the evolution of RTI-driven turbulence, and suggested that it is more likely that DDT would occur at somewhat higher densities, than the canonical value of $2\times10^7$ g cm$^{-3}$ \citep{Hoeflich1995}.

\begin{figure*}
    \begin{center}
    \includegraphics[width=\linewidth]{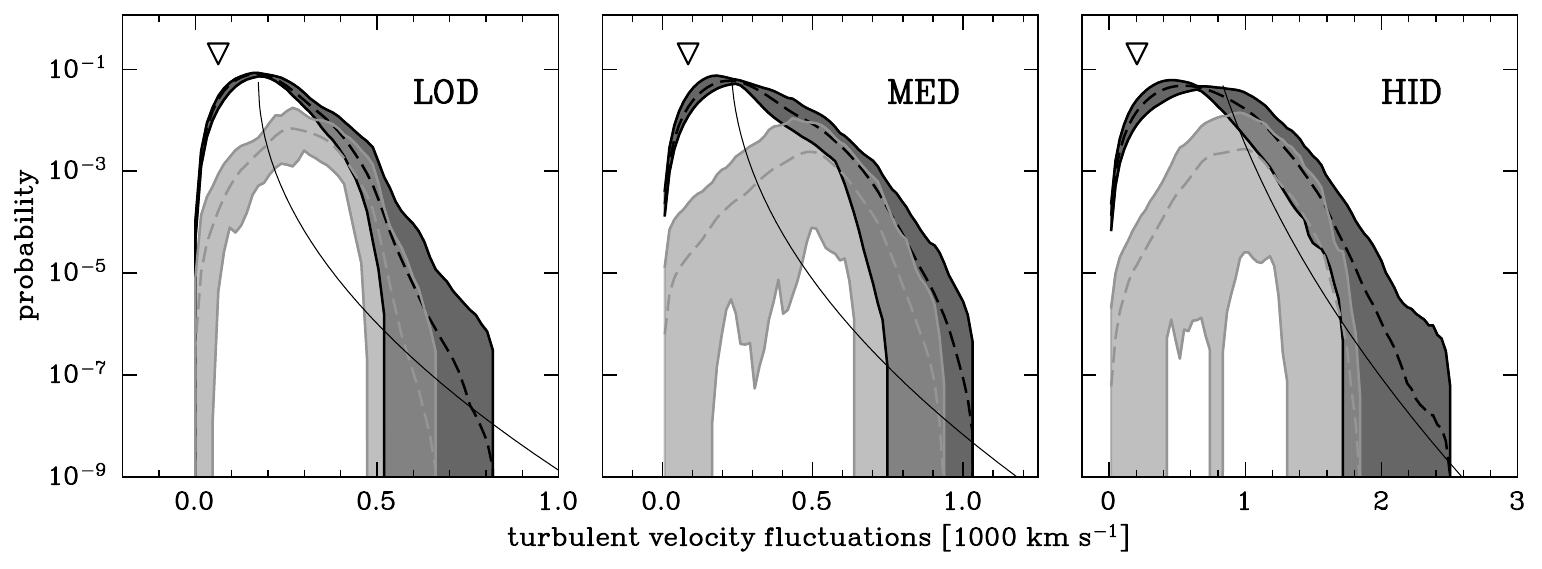}
    \caption{Time-averaged PDFs of turbulent velocity fluctuations for the material in the TPF region for the LOD (left panel), MED (middle panel), and HID (right panel) model. Time-averaged velocity PDFs marginalized in terms of the flame progress variable are plotted with dark gray and light gray dashed lines for all material ($\phi \le 1$) and pure fuel ($\phi \le 1\times 10^{-5}$), respectively. The associated gray-shaded bands depict the extreme ranges of the turbulent velocity fluctuations for the respective plasma component within the analyzed time-series. The inverted triangle symbols mark the turbulence intensity at the scale of $\approx 8$ km. Thin solid lines correspond to approximate log-normal distributions using both Eq.\ 1 and the fit parameters given in Table 2 by \citet{Roepke07} for $t=0.080$ seconds. The original values of parameters given by \citet{Roepke07} were adjusted to scale up theoretical distributions to approximately match our data. See text for discussion.}
    \label{f:velpdf}
    \end{center}
\end{figure*}

Our results do not support existence of high velocity fluctuations assumed by the above authors. Analysis of our simulation results indicates the velocity fluctuations on the scale of $\approx 8$ km, corresponding to $k=3$ (cf.\ Fig.\ \ref{f:ke}), are about 63, 86, and 205 km/s in LOD, MED, and HID models, respectively. Those estimated velocities are indicated by triangle symbols in Fig.\ \ref{f:velpdf}. The velocity fluctuations seem to follow a log-normal distribution rather closely at lower densities (LOD and MED models), while a better agreement with the postulated by \cite{Roepke07} exponential distribution can be found in our highest density model (HID model). In general, turbulent motions in the fuel (depicted with light gray dashed line in Fig.\ \ref{f:velpdf}) are characterized by steeper, and possibly closer to log-normal, distributions in contrast to flatter distribution of velocities in the total velocity field (indicated with dark gray dashed line in Fig.\ \ref{f:velpdf}). We conclude that there are significant qualitative and quantitative differences in turbulent velocity distributions between our models and those analyzed by \cite{Roepke07} and assumed by later authors (e.g., \cite{Aspden2010, Schmidt+10}).

Motivated by the numerical results of \cite{Roepke07}, in particular the reported deviations away from the expected log-normal distribution of the velocity fluctuations, \cite{Schmidt+10} extended the intermittency model originally developed by \cite{Pan+08} to the stirred-flame regime \citep{Woosley+09}, and used it to characterize the dependence of flame-turbulence interactions on the physical scale. Their model provides probability estimates of DDT occurrence for the given level of velocity fluctuations, expressed in terms of the energy dissipation rate, and the Karlowitz number, which quantifies relative impact of turbulence on the laminar flame. Their model predicts high probability of DDT in the case of low velocity fluctuations on ZDDT scales, and excludes ZDDT if the velocity fluctuations exceed 500 km/s, as required by \cite{Woosley+09}. We note that the low velocity fluctuations favored by the \cite{Schmidt+10} model are observed in our models. Their result was obtained assuming a log-normal velocity fluctuation distribution consistent with the velocity distributions found in this work.
%
%
In parallel to the above investigations, Woosley and his collaborators investigated the evolution of thermonuclear flames on the microphysical scale \citep{Bell2004, Zingale+05, Aspden2008} and using parameterized turbulence models at RTI-dominant scales \citep{Woosley+07, Woosley+09, Aspden2010}. The two earlier studies by this group concerned with flame evolution on small scales could offer important information in the context of this work in part related to the fuel preconditioning, provided the evolution was studied on sufficiently long timescales and on sufficiently large domains to quantify the thermal expansion of the turbulent region directly associated with the flame front. On the other hand, the results obtained by this group on scales comparable to our model, in particular work by \cite{Aspden2010}, resulted in conclusion that transition to detonation is most likely when the flame filling factor is high and turbulent intensities reach at least 20\% of the sound speed. 

Regarding the first condition put forward by these authors, we do not observe any significant level of flame packing in the TPF region. This statement is limited by numerical resolution of our model, combined with the numerical thickness of the flame profile. Combination of those two factors effectively constrains the physical scale on which we can observe flame-flame interactions to multiple times the mesh resolution ($\approx$62 m). 

For the second condition, average turbulence Mach numbers of the TPF fuel range between about 0.1 and 0.2, with the higher value, required by \cite{Aspden2010}, being realized only in the HID model. Furthermore, the extreme fuel velocities observed in the TPF region, indicated by light-gray solid lines in Fig.\ \ref{f:velpdf}, reach Mach numbers of about 0.18, 0.20, and 0.36 in LOD, MED, and HID models, respectively. These velocity excursions are only by a factor $\lesssim 2$ above the average, which is far lower than the level of excursions by a factor of 5 reported by \cite{Aspden2010}. This is despite the fact that our models retain compressibility effects, which comprise about 20\% of the velocity field, and are expected to produce a high-velocity tail as shown in Fig.\ \ref{f:velpdf}, but which were not accounted for in the low Mach number approach employed by \cite{Aspden2010}. In passing we note that this discrepancy is identified based on the relative change of velocities, distribution of which is only indirectly represented through a turbulent Damk\"ohler number, as defined by Eq.\ 3 in \citet{Aspden2010}. 

Nevertheless, the issue of the missing high end turbulent velocity distribution tails in our models remains. One possible reason for this qualitative difference between the model velocity distributions is that turbulence is naturally driven in our simulations by buoyancy combined with RTI, while \cite{Aspden2010} used parameterized spectral-driving. Even though in their approach the drive is somewhat reduced in the ash, this drive component is entirely missing in our simulations, consistent with SN Ia physics. That this difference in the prescribed computational models might indeed be responsible for the inferred difference in the turbulent velocity distributions follows from the observation that, as indicated in Fig.\ \ref{f:velpdf}, the highest observed velocities in our model are associated with at least partially burned material, rather than the fuel. (In the figure, the dark-gray filled contours corresponding all material in the TPF region are extending towards higher velocities beyond fuel, marked with the light-gray filled contours.) This implies that even though the driving is no longer present in our RTI-driven models, it is the burned material where the highest velocities are observed. It is therefore conceivable that injecting additional kinetic energy into this material will further increase its velocities and amplify velocity excursions possibly to reach the level reported by \cite{Aspden2010}.

Finally, in the context of RTI-driven turbulence, \cite{Woosley+11} considered the possibility of ZDDT in oxygen flames. One expects that this type of flame emerges deeper in the RTI flame brush given that the burning timescale for oxygen is substantially longer than for carbon \citep{Khokhlov1991a, Calder2007}. \cite{Woosley+11} theorized, based on one-dimensional parameterized turbulent flame models, that oxygen may undergo supersonic burning provided that sufficiently large velocity fluctuations exist on sufficiently large scales behind the carbon flame. Although we do observe oxygen burning in our models, our analysis is limited to the TPF region located above the oxygen flame, and therefore we can only provide indirect evidence in the context of the \cite{Woosley+11} scenario. Given that, as discussed earlier, our RTI-based turbulent velocity fluctuation estimates are significantly lower than the 500 km/s on the scale of 10 km required by \cite{Woosley+11}, we conclude that the likelihood of ZDDT is much smaller than considered by \cite{Woosley+11}. In passing we would like to note that our spectrally-driven turbulence models \citep[see, for example,][]{Brooker2021} do produce several kilometer wide regions with temperature perturbations on the order of 10\%-20\% before carbon ignites (B. Gusto, private communication).
\subsubsection{Turbulence-induced DDT}
\cite{Poludnenko2019} proposed a unified theory of turbulently-induced DDT (tDDT) connecting known phenomena in chemical combustion to thermonuclear explosions in SNe Ia. They demonstrated that if the flow is sufficiently energized (by either turbulence or a shock), burning rate in the perturbed flow may reach levels sufficient to form and support a sustained detonation wave. In their first SN Ia application, unsteady burning of the thermonuclear flame was demonstrated to cause a pressure buildup and localized generation of weak shocks. In their second SN Ia application, those shocks were shown capable of producing detonations as they are gradually strengthened in the process of interacting with a turbulent flame. 

Of note is that the proposed by \cite{Poludnenko2019} tDDT mechanism can operate in the regime of a relatively loosely packed turbulent flame, where the laminar flame sheet is folded while the internal flame structure is largely unaffected by turbulence. Therefore, the tDDT theory is qualitatively different from the previously reviewed SN Ia DDT theories that required strong turbulence-flame interaction. These distributed flame DDT models require turbulent intensities on the order of 500 km/s at scales of several kilometers for densities close to $1\times 10^7$ g cm$^{-3}$, which may not be achievable in realistic SN Ia explosion models and were not realized in our relatively well-resolved simulations. 

\cite{Poludnenko2019} further elaborates on the idea of laminar flame packing that effectively increases the flame surface density for a given volume. They show that a characteristic flame volume density exists at which CJ conditions are met. A lower bound on the flame volume density for their tDDT mechanism to occur corresponds to the maximally packed flame. In that regime, flame sheet separations are on the same scale as the laminar flame thickness. Following the common assumption of the turbulent intensity on the order of 100 km/s at 10 km scale, Poludnenko et al.\ estimated that the greatest probability of tDDT is associated with the density of about $3\times10^7$ g cm$^{-3}$ and the characteristic physical scale, corresponding to the Chapman-Jouget length, of about $2\times10^3$ cm. We note that this scale is by a factor of about 30 smaller than the mesh resolution in our models, which poses a challenge for modelers motivated in obtaining self-consistent picture of the RTI-driven flame evolution below gravity-dominated scales.

The tDDT model assumes the flow field can be represented by homogeneous and isotropic Kolmogorov-type turbulence in quasi-steady state. These assumptions should be contrasted with our findings that indicate that the RTI-driven turbulence is far from steady, even without accounting for possible additional large-scale perturbations present in explosion models. The turbulent intensities vary significantly between the fuel and ash, and redistribution of the buoyant energy that drives turbulence does not produce isotropic velocity fluctuations even in regions deep behind the leading edge of the flame where the remaining amount of driving (fuel) is small.

Furthermore, tDDT conditions are estimated using turbulent intensities found in relatively poorly resolved, centrally-ignited pure deflagration SN Ia explosion models by \cite{Khokhlov1995} and \cite{Roepke07}. Our simulations provide improved information about turbulent intensities on ZDDT scales, which differ from the original estimates. On the lower end of the density regime of interest ($\approx 1.4\times 10^7$ g cm$^{-3}$), our MED model points to somewhat lower intensities (86 km/s) on ZDDT scales than typically assumed (100 km/s), while on the higher end of densities explored ($\approx 6.3\times 10^7$ g/cc) our HID model points to higher intensities at 205 km/s. Given these improved estimates, one may expect the most likely tDDT density to be higher than the estimated by Poludnenko et al.\ value of $3\times 10^7$ g/cc.
\section{Summary and conclusions}\label{s:conclusions}
We have performed a series of computer simulations of the evolution of RTI-driven thermonuclear deflagrations in a centrally ignited massive white dwarf scenario. Initial conditions were obtained from 2D pure deflagration simulations of a realistic, Chandrasekhar mass progenitor, at select times when the flame enters outer layers of the progenitor and delayed-detonation SN Ia explosion models commonly assume DDT takes place. These models show strong promise of possibly explaining a large number of SN Ia events, but the DDT mechanism has not been demonstrated in a self-consistent manner. This work quantifies conditions that exist deep inside the flame brush where turbulence-flame interactions are strongest and occurrence of DDT is presumably the most probable.

For the conditions expected when a centrally ignited deflagration front reaches the outer layers of a Chandrasekhar mass white dwarf progenitor with fuel densities $\rho_\mathrm{f}\approx5{\times}10^6{\relbar}6\times 10^7$ g cm$^{-3}$, the main results of our work can be summarized as follows:
\begin{enumerate}
    \item We find for the first time evidence for strong flame-fuel interaction, which results in shortening fuel ignition times in layers close to the flame front by about two to five orders of magnitude, with stronger shortening at higher densities. We attribute this effect to adiabatic compression of fuel due to the flame-induced buoyant acceleration.
    \item Depending on the initial state of the fuel, adiabatic heating may result in fuel burning ahead of the flame front, including ignition and possibly transition to detonation, if the fuel layer is suitably preconditioned. This effect should be included in the next generation of RTI-driven deflagrations.
    \item We were unable to reproduce preconditioning required by ZDDT found in our earlier SD turbulence models (BFP21) despite similar fuel conditions and energy drive. This discrepancy between model outcomes can be due to differences in assumed physics with RTI drive being highly time-dependent, RTI forcing limited to a subset of flame front segments, and self-heating effects not accounted for in this work. Preconditioning mechanism identified in SD models requires long-wavelength compressive modes, which are largely absent under RTI forcing.
    \item However, we find evidence for preconditioning of fuel in RTI spikes, which closely resemble fuel channels found in SD models. Contribution of long compressive modes to fuel preconditioning in spikes is also possible in SN Ia explosion models, which do not restrict turbulent forcing spectral range. 
    \item The turbulent velocity fluctuations on scales of several kilometers and at fuel densities commonly considered relevant to the ZDDT mechanism increase from about 60 km/s to about 200 km/s as the density increases from about $5\times10^6$ g cm$^{-3}$ to about $6\times 10^7$ g cm$^{-3}$. These turbulent intensities are substantially lower than the estimated 1000 km/s based on early explosion models \citep{Khokhlov1995,Roepke07} and as typically required for DDT models operating in the distributed regime of burning \citep{Woosley+07,Schmidt+10}.
    \item The turbulence that exists deep in the flame brush region, identified in this work as the TPF region, has a strongly intermittent character. This is evidenced by the saturation of velocity structure function scaling exponents in both longitudinal and lateral directions, which do not significantly change in value beyond the third order structure function. This behavior is in strong contrast to scalings predicted by models that assume extended self-similarity, such as \cite{She1994}, and as found by \cite{CiaraldiSchoolmann2009} in the explosion model by \cite{Roepke07}.
    \item We attribute the high degree of intermittency to first variable RTI injection rate, which varies by a modest 50\% at the lowest density model but experience variation by a factor of about 3 at the high end of the considered density range. Furthermore, the energy injection timescale is several times longer than the TPF eddy turnover time. Strong saturation of the velocity structure function exponents is observed at most times, however, velocity fluctuations occasionally follow scalings close to those predicted by models that assume extended self-similarity.
    \item The turbulent velocity distributions found in our models show a fair agreement with the exponential Ansatz used to describe velocity fluctuations in low resolution explosion models \citep{Roepke07}. However, the strongest deviations from lognormality are only observed in the burned material; the distributions of fuel velocities appear systematically closer to log-normal than those of the ash velocities. Furthermore, the velocity distributions in both fuel and ash become closer to log-normal as the density decreases. This might be indicative of the flow becoming gradually more correlated on small scales at lower densities, which might be a sign of possible transition to a distributed burning regime.
    \item On the driving scale, the flame surface fractal dimension exceeds 2, and gradually decreases through the flame brush as the flame fragments and buoyancy weakens. The observed values of fractal dimension are generally lower than those found by \cite{CiaraldiSchoolmann2013} in the \cite{Roepke07} explosion model. The lower flame fractal dimension implies lower estimated probability of DDT in models requiring high intensity of turbulence on the ZDDT scales.
    \item In the context of a recently proposed tDDT model \citep{Poludnenko2019}, the lower degree of flame packing suggested by our data will result in less intense flame-shock interactions assumed in that model. Additionally, the tDDT shock generation mechanism might be affected by a modified thermal fuel structure due to the observed intense flame-fuel interactions. Better resolved numerical simulations with self-consistent fuel state are needed to assess the effect of these factors on tDDT model.
\end{enumerate}
Although the flame-in-a-box approach used in this work closely represents conditions found in SN Ia explosion models on buoyancy-mediated scales, it has its limitations. During the explosion, the RTI-unstable flame generates perturbations on scales greater than the largest driving scale available in models executed on domains with limited lateral extent. Restricted geometry also sets a forcing timescale on the order of the large-scale eddy turnover time, while in explosion models one can expect a certain degree of coupling between different large-scale RTI modes.

The model limitations are primarily due to the very high computational costs, in particular the required computing time, which necessitates a compromise between model resolution and the size of the computational box. One important model improvement that we mentioned, and which is not expected to substantially increase computational cost, is inclusion of fuel-restricted nuclear self-heating to account for possible energy release in adiabatically heated fuel. With the system's behavior now examined for a relatively broad range of initial conditions, one could explore evolution for one select density, such as the estimated most probable tDDT density.

\section*{Acknowledgments}
This research used resources of the National Energy Research Scientific Computing Center, which is supported by the Office of Science of the U.S.\ Department of Energy under Contract No.\ DE-AC02-05CH11231. The software used in this work was in part developed by the DOE Flash Center at the University of Chicago. Data visualization was performed in part using VisIt \citep{HPV:VisIt}. This research has made use of NASA's Astrophysics Data System Bibliographic Services.
\section*{Data availability}
The data underlying this article will be shared on reasonable request to the corresponding author.
%
%
%
%

%
%
\bibliographystyle{mnras}
\bibliography{references}
%
%
%
\label{lastpage}
\end{document}